\documentclass[a4paper,pra,superscriptaddress,twocolumn,nofootinbib,floatfix,accepted=2025-12-21]{quantumarticle}
\pdfoutput=1

\usepackage{color}
\usepackage{graphicx}
\usepackage{float}
\usepackage{dcolumn}
\usepackage{braket}
\usepackage{physics}
\usepackage{amsmath,amssymb}
\newcommand{\figuref}[1]{\mbox{Figure~\ref{#1}}}
\usepackage[breaklinks=true,colorlinks,citecolor=blue,linkcolor=blue,urlcolor=blue]{hyperref}
\usepackage{makeidx}
\usepackage[capitalise]{cleveref}

\usepackage[dvipsnames]{xcolor}

\begin{document}

\author{Daniele Lamberto}
\orcid{0009-0007-4716-5661}
\email{daniele.lambe@gmail.com}
\affiliation{Dipartimento di Scienze Matematiche e Informatiche, Scienze Fisiche e  Scienze della Terra, Universit\`{a} di Messina, I-98166 Messina, Italy}

\author{Gabriele Orlando}
\orcid{0009-0006-7683-4505}
\affiliation{Dipartimento di Scienze Matematiche e Informatiche, Scienze Fisiche e  Scienze della Terra, Universit\`{a} di Messina, I-98166 Messina, Italy}

\author{Salvatore Savasta}
\orcid{0000-0002-9253-3597}
\affiliation{Dipartimento di Scienze Matematiche e Informatiche, Scienze Fisiche e  Scienze della Terra,	Universit\`{a} di Messina, I-98166 Messina, Italy}

\newcommand{\figref}[1]{\mbox{Fig.~\ref{#1}}}
\newcommand{\figpanel}[2]{Fig.~\hyperref[#1]{\ref*{#1}(#2)}}
\newcommand{\figurepanel}[2]{Figure~\hyperref[#1]{\ref*{#1}(#2)}}
\newcommand{\figpanels}[3]{Figs.~\hyperref[#1]{\ref*{#1}(#2)-(#3)}}

\newcommand{\be}{\begin{equation}}
\newcommand{\ee}{\end{equation}}
\newcommand{\bea}{\begin{eqnarray}}
\newcommand{\eea}{\end{eqnarray}}

\renewcommand{\eqref}[1]{\mbox{Eq.~(\ref{#1})}}
\newcommand{\eqaref}[1]{\mbox{Equation~(\ref{#1})}}

\title{Superradiant Quantum Phase Transition in Open Systems: System-Baths Interaction at the Critical Point}

\maketitle

\begin{abstract}

The occurrence of a second-order quantum phase transition in the Dicke model is a well-established feature. 
On the contrary, a comprehensive understanding of the corresponding open system, particularly in the proximity of the critical point, remains elusive.
When approaching the critical point, the system inevitably enters first the system-bath ultrastrong coupling regime and finally the deep-strong coupling regime, causing the failure of usual approximations adopted to describe open quantum systems. 
In the thermodynamic limit, we study the interaction between the Dicke model and bosonic bath fields without resorting to additional approximations, which usually relies on the weakness of the system-bath coupling.
We find that the critical point is not affected by interactions with environments displaying metastable minima. Moreover, such interactions cannot affect the system ground-state condensates in the superradiant phase, whereas the bath fields are {\em infected} by the system and acquire macroscopic occupations.
The obtained reflection spectra display lineshapes which become increasingly asymmetric, both in the normal and superradiant phases, when approaching the critical point.
\end{abstract}

\section{Introduction}

The Dicke model was one of the first models historically introduced in the study of the light-matter interaction \cite{Dicke54, Hepp_Lieb1973, Wang1973}. It was originally proposed to describe the interaction of a single cavity mode with a collection of $N$ atomic dipoles, treated as two-level systems (TLSs).
This model exhibits a classical second-order equilibrium phase transition when the coupling strength exceeds a critical value, entering a phase in which the collective atomic polarization and the photonic field are finite even in the absence of external driving. This widely studied phenomenon \cite{kirton2019introduction, garraway2011dicke} is known as superradiant phase transition (SPT). The Dicke model Hamiltonian also exhibits a quantum phase transition (QPT) \cite{sachdev1999quantum}, which can occur at zero temperature \cite{NatafCiuti2010, Brandes03, Brandes03_PRL} by tuning the light-matter coupling across a quantum critical point, which is also referred to as ground-state photon condensation \cite{Andolina2019, Zueco2021}.
However, according to a number of no-go theorems, the SPT (and also the corresponding QPT \cite{NatafCiuti2010, Knight1978, Lamberto2024}) is forbidden for systems interacting with photons by only their electric polarization \cite{NatafCiuti2010, Andolina2019, Knight1978, Andolina2022}. In the Coulomb gauge, it is the diamagnetic term in the Hamiltonian, neglected in the Dicke model, which prevents the SPT \cite{Knight1978, Zakowicz1975}.

This equilibrium SPT has been attracting enduring attention since several decades, but its experimental demonstration still remains very challenging.
Numerous non-equilibrium realizations of SPTs have been proposed over the years \cite{Dimer2007}.
Implementations of effective Dicke Hamiltonians in driven-dissipative systems have been reported, e.g., in cold-atom systems driven by laser fields \cite{Baumann2010dicke, Nagy2010, Zhiqiang2017nonequilibrium, kirton-keeling2018superradiant, Nagy2015, Nagy2016} and trapped ions \cite{Genway2014}. However, despite strong analogies with the equilibrium SPT, non-equilibrium SPTs in driven-dissipative systems are inherently different phenomena \cite{kirton2019introduction, Corps2023, Heyl2015, Heyl2018, Zvyagin2016}. 

Theoretical proposals for the observation of equilibrium SPTs consider circuit QED systems \cite{NatafCiuti2010, Debernardis2018, Viehmann2011, Lambert2016, Bamba2016, forn2017ultrastrong, Ashhab2019}, electron gases that either display a Rashba spin-orbit coupling \cite{Nataf2019} or interact with a spatially varying electromagnetic field \cite{Nataf2019, Polini2020, Guerci2020}, magnetic molecules that couple to superconducting microwave resonators via the Zeeman interaction \cite{Zueco2021,Mercurio2024, Ghirri2023}. 
Interesting related applications involve quantum computing in models predicting a QPT \cite{Albash2018RMP, Bernien2017probing, Bohnet2016trappedions}.
A recent proposal \cite{bamba2022magnonic} considering Er$^{3+}$ spins cooperatively interacting with a magnonic field, playing the role of photons, has led to a spectroscopic evidence for an equilibrium SPT \cite{kono2024observation}.


The SPT  has been exhaustively studied theoretically in the isolated system, both at zero and nonzero temperature \cite{Hepp_Lieb1973, Brandes03, Brandes03_PRL}. 
It has been shown that the system's ground state displays a two-mode quantum squeezed vacuum, reaching perfect squeezing at the SPT critical point \cite{hayashida2023perfect}.
Moreover, the ground-state entanglement between the atoms and the field diverges logarithmically at the critical coupling for $N \to \infty$ \cite{Lambert2004}. 
On the other hand, no clear understanding of the open-system counterpart has been reached yet in the proximity of the QPT critical point, especially when considering equilibrium situations. Given the recent  progress \cite{kono2024observation}, which reported  the first (to our knowledge) long-sought experimental observation of an equilibrium SPT at low temperatures, a fully rigorous quantum treatment of the open Dicke model is required. Moreover, an accurate description of decoherence and of the relationship between the system and the input-output fields near critical points is essential for the development of criticality-enhanced quantum sensing \cite{Chu2021,Ilias2022,hotter2024}.

A key feature of the equilibrium SPT in the thermodynamic limit is the softening of the lowest polariton mode, which vanishes at the critical point, signaling that the transition to a photon condensate state is a second-order quantum phase transition.  This makes the standard treatments impossible, given that the usual approximations employed when studying open quantum systems fail, as they are based on the smallness of the ratio between the loss rate and the relevant resonance frequencies of the system $\gamma/\omega$. Indeed, sufficiently close to the critical point, the system inevitably enters first the system-bath ultrastrong coupling regime  ($\gamma/\omega > 0.1$) and finally the deep-strong coupling regime ($\gamma/\omega > 1$). In the last decade, light-matter systems in the ultrastrong and deep-strong coupling regimes have been widely studied both theoretically and experimentally \cite{frisk2019ultrastrong, forn-diaz_review}. It has been shown that these regimes require a particular care in the description of both decoherence and input-output fields \cite{Beaudoin2011, Ridolfo2012, Settineri2018, Mercurio2023}, since standard approximations based on the rotating-wave-approximation fail. However these open quantum systems are usually studied under the assumption of weak system-bath interactions. Very recently, a general linear response theory for materials collectively coupled to a cavity, including symmetry-broken phases, has been presented \cite{Roche2025}. 
Within this approach, however, the system-bath interaction is not taken into account in the Hamiltonian description, and the spectroscopic response of the cavity is calculated by analytic continuation from the system Green's function.

Several questions remain open: (i)  What is the impact of these system-bath extreme interaction regimes on the QPT? (ii) Assuming that the environment does not destroy the QPT, how does it affect the critical point and the ground state condensates? (iii)  Are the external thermal baths affected by the QPT? (iv) Can the nonclassical properties of the system ground state \cite{hayashida2023perfect, Lambert2004} be observed by detection of vacuum fluctuations \cite{Riek2015direct, Benea2019electric, Lindel}? (v) How to calculate the (observed \cite{kono2024observation}) spectroscopic features in proximity of the critical point? (vi) How do the answers to the above questions depend on the specific spectral densities of the thermal baths?

In this paper, we present a full quantum description of the open Dicke model in the thermodynamic limit, based on quantum Langevin equations, not restricted by the smallness of $\gamma / \omega$ and that can be applied for essentially any range of parameters, differently from approaches such as the Lindblad master equation that rely on the rotating wave approximation (RWA) or the weakness of the system-bath interaction.
The framework presented here is able to accurately describe the equilibrium open Dicke model in the proximity of the mode softening.
Within this approach, the superradiant phase of the open Dicke model is addressed by diagonalizing the full Hamiltonian describing the open quantum system. 
We find that a large class of thermal baths consisting of an infinite number of harmonic oscillators (normal modes) interacting with the system via a potential displaying a metastable minimum and with {\em well-behaved} densities of states (properly vanishing as $\omega \to 0$) do not affect the critical point of the corresponding closed system. This result is in agreement with recent experimental data \cite{kono2024observation}. We will also show that, in this case, the ground-state condensation occurring in the system influences the bath state, but not vice versa. 
In contrast, previous theoretical studies of driven-dissipative systems, based on either the master-equation approach \cite{Brandes_openDicke2013, FredrikLambert2024, DallaTorre2013} or the Keldysh path-integral formalism \cite{Nagy2015, Nagy2016}, predict a quantum critical point that depends on the cavity dissipation rate. This discrepancy is expected, given the fundamentally different physical settings and theoretical frameworks employed. Indeed, in those works, approximations such as the RWA for the system-environment coupling are employed, which is appropriate because the system dynamics is described in the high-frequency rotating frame imposed by the external drive. In our case, however, these approximations are not valid, as our treatment explicitly addresses the equilibrium regime without considering any rotating frame.

Furthermore, our theoretical approach allows us to calculate the spectral properties of the system when probed by a coherent weak tone at any value of the critical parameters, for different bath spectral densities, and for arbitrary large damping rates.
This framework is not limited to systems described by the Dicke Hamiltonian, but can also be applied to characterize similar spectral behaviors observed near a quantum critical point \cite{Libersky2021observationQPT}.

The structure of this paper is as follows. 
In Sec.~\ref{sec:standard_Dicke}, we outline the main features of the standard Dicke model for a closed system and establish the notation used throughout the manuscript.
In Sec.~\ref{sec:open_dicke}, we investigate the open Dicke model through our approach, based on the quantum Langevin equations. Specifically, in Secs.~\ref{sec:normal_phase} and \ref{sec:superradiant_phase}, we derive the properties of the open system in the normal and superradiant phases, respectively, while Sec.~\ref{sec:excitation_energies} provides a detailed discussion of these results. 
The theoretical framework for calculating coherent spectra, applicable to arbitrary bath density of states, is developed in Sec.~\ref{sec:theory_spectra}, where we also present several coherent reflection spectra as illustrative examples.
In Sec.~\ref{sec:qX}, we explore alternative system-bath interactions and their implications on the system properties and the associated SPT.
A discussion on the squeezing properties of the Dicke model, including their characterization and observability in an open system, is presented in Sec.~\ref{sec:squeezing}.

\begin{figure}[t]
    \centering
    \includegraphics[width=\linewidth, scale=0.2]{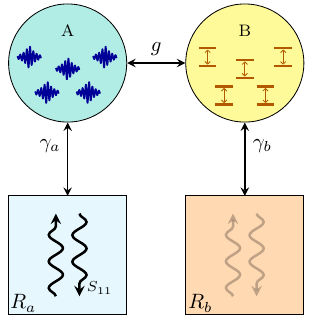}
    \caption{\textbf{Schematic sketch of the system.} Two interacting subsystems, A and B, with coupling strength $g$. Each component interacts with its own thermal bath. The interaction of a subsystem with its own reservoir can also be regarded as an input-output port, through which the system can be excited and probed. In this work we will consider explicitly a single tone coherent excitation of subsystem A and calculate the corresponding reflection coefficient $S_{11}$. This framework can be easily generalized to include the interaction with additional thermal baths.}
    \label{fig:ref_system}
\end{figure}

\section{Standard Dicke Model and superradiant phase transition} \label{sec:standard_Dicke}

In this section, we briefly present the main features of the Dicke model for an isolated system at zero temperature, \textit{i.e.}, without considering its interaction with the external environment, which will be discussed in the Section \ref{sec:open_dicke}. The Dicke Hamiltonian, which describes a single bosonic mode with resonance frequency $\omega_a$ interacting with a collection of $N$ identical TLSs with transition frequency $\omega_b$ (see \figref{fig:ref_system}), is usually written as
\begin{equation} \label{eq:Dicke_Ham}
    H_{\rm sys} = \hbar \omega_a a^\dag a + \hbar \frac{\omega_b}{2} \sum_{i=1}^N \sigma^z_i + \hbar \frac{g}{\sqrt{N}} \left( a^\dag + a \right) \sum_{i=1}^N \sigma^x_i \, ,
\end{equation}
where $a$ is the bosonic annihilation operator, $\sigma^\alpha_i$ (with $\alpha = x, y, z$) are the usual Pauli operators associated to the $i$-th TLS and $g$ is the coupling strength.
This model does not include the so-called $A^2$ (or $P^2$) term, which must be included when describing electric dipoles coupled to the electromagnetic field of a cavity in the Coulomb (dipole) gauge \cite{Lamberto2024, Cohen_book, Garziano2020}. However, it is well-suited for systems displaying a Zeeman-like interaction \cite{bamba2022magnonic}. A further generalization of the Dicke model allows for non-homogeneous coupling between the two-level systems (TLSs) and the radiation field \cite{kirton2019introduction, Hioe73}.
Equation (\ref{eq:Dicke_Ham}) can be rewritten by introducing collective spin operators $J^\alpha = \sum_i \sigma^\alpha_i / 2$, satisfying the usual angular momentum algebra.
Henceforth, we focus our study to the manifold of maximum angular momentum, which is composed by the states that maximally couple with the bosonic field \cite{Brandes03}. This is equivalent to fix $j = N/2$, with $j$ being the eigenvalue of the total angular momentum operator ${\bf J}^2$. 
Indeed, at zero temperature and large $N$, the ground and the first excited states reside within this manifold, making it the most relevant for investigation.
We now introduce the Holstein-Primakoff (HP) transformations \cite{HolsteinPrimakoff, Kapor_HolsteinPrimakoff1991, GrossoPastori_book}, defined by $J^z = \hbar \left( b^\dag b - N/2 \right) , \, J^+ = \hbar b^\dag \sqrt{N - b^\dag b}$ and $J^+ = (J^-)^\dag$, which yields to the bosonic version of the Dicke Hamiltonian
\begin{equation} \label{eq:Dicke_Ham_boson}
    \begin{aligned}
        H_{\rm sys} = & \,\, \hbar \omega_a a^\dag a + \hbar \omega_b b^\dag b \\ 
    & + \hbar g \left( a^\dag + a \right) \left( b^\dag \sqrt{1 - \frac{b^\dag b}{N}} + \sqrt{1 - \frac{b^\dag b}{N}} \;  b \right) \, .
    \end{aligned}
\end{equation}

\begin{figure}[t]
    \centering
    \includegraphics[width=\linewidth]{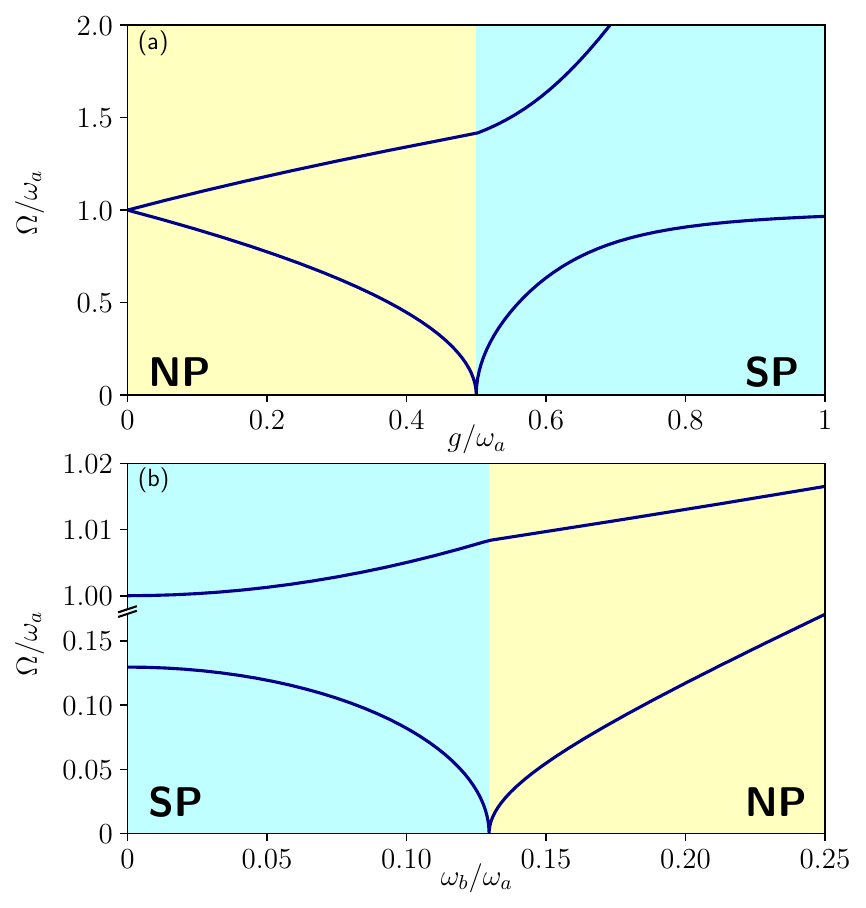}
    \caption{{\bf Excitation energies of the isolated system.} Upper and lower excitation energies for the isolated system in the normal (yellow background) and superradiant (cyan background) phases as a function of the {\bf (a)} normalized coupling and {\bf (b)} frequency ratio. Parameters: {\bf (a)} $\omega_b / \omega_a = 1$, {\bf (b)} $g / \omega_a = 0.18$.}
    \label{fig:disp_rel_closed}
\end{figure}

In the thermodynamic limit (corresponding to $N \to \infty$, while $\rho = N/V$ remains finite), the Dicke Hamiltonian predicts a second-order QPT for a critical coupling $g_c = \sqrt{\omega_a \omega_b} / 2$ \cite{NatafCiuti2010, Brandes03}. 
While in the so-called \textit{normal phase} ($g < g_c$) the ground state of the system does not exhibit any macroscopic condensate, in the \textit{superradiant phase} ($g > g_c$) both fields acquire a non-zero ground-state coherent occupation, which structurally change the properties of the system.
In the normal phase, the lowering and raising operators in the HP maps can be linearized in the low-excitation regime, leading to the effective Hamiltonian
\begin{equation} \label{eq:H_normal}
    H_{\rm NP} = \hbar \omega_a a^\dag a + \hbar \omega_b b^\dag b + \hbar g \left( a^\dag + a \right) \left( b^\dag + b \right) \, .
\end{equation}
Such Hamiltonian can be diagonalized by considering the corresponding Hopfield-Bogoliubov matrix $\bf A$, whose eigenvalues represent the excitation energies of the system, $\Omega$. Such matrix in the normal phase, ${\bf A}_{\rm NP}$, reads
\begin{equation}
    {\bf A}_{\rm NP} =
    \begin{pmatrix}
        \omega_a & 0 & g & g \\
        0 & -\omega_a & -g & -g \\
        g & g & \omega_b & 0 \\
        -g & -g & 0 & -\omega_b
    \end{pmatrix}\,.
\end{equation}
The lowest excitation energy $\Omega_-$ vanishes as $g \to g_c$ as expected (see \figref{fig:disp_rel_closed}(a)), thus signaling the presence of the SPT. On the other hand, in the superradiant phase obtained for $g > g_c$, the effective Hamiltonian in \eqref{eq:H_normal} is no longer valid since both fields acquire a macroscopic coherent occupation. To take into account these condensations, we shift the bosonic operators as $a = a_{\rm s} + \sqrt{\alpha} $ and $ b = b_{\rm s} - \sqrt{\beta}$,
where $\alpha$ and $\beta$ are c-number of order $O(N)$, while $a_{\rm s}$ and $b_{\rm s}$ are bosonic operators describing the fluctuations with respect to the respective mean value. By inserting these definitions into \eqref{eq:Dicke_Ham_boson} and imposing the equilibrium condition in the resulting expression (which corresponds to the vanishing of the linear terms in these bosonic operators), we obtain non-zero macroscopic mode occupations in the superradiant phase, which are given by

\begin{equation}\label{eq:h_mini}
    \alpha = \frac{N g^2}{\omega_a^2} \left( 1 - \frac{1}{\lambda^2} \right) \, , \quad \beta = \frac{N}{2} \left( 1 - \frac{1}{\lambda} \right) \, ,
\end{equation}
where $\lambda = 4 g^2 / \omega_a \omega_b = g^2 / g_c^2$. The corresponding effective Hamiltonian in the superradiant phase is
\begin{equation}\label{eq:H_sr}
    H_{\rm SP} = \omega_a a_{\rm s}^\dag a_{\rm s} + \tilde{\omega}_b b_{\rm s}^\dag b_{\rm s} +  \tilde{g} ( a_{\rm s} + a_{\rm s}^\dag) ( b_{\rm s} + b_{\rm s}^\dag ) + D ( b_{\rm s} + b_{\rm s}^\dag )^2,
\end{equation}
where $\tilde{\omega}_b = \omega_b \left( \lambda + 1 \right)/2$, $\tilde{g} = g_c \sqrt{2/ (\lambda + 1)}$ and $D = \omega_b ( \lambda - 1 ) ( 3 \lambda + 1 ) /8 ( \lambda + 1 )$.
The Hopfield-Bogoliubov matrix in the superradiant phase reads
\begin{equation}
    {\bf A}_{\rm SP} =
    \begin{pmatrix}
        \omega_a & 0 & \tilde{g} & \tilde{g} \\
        0 & -\omega_a & -\tilde{g} & -\tilde{g} \\
        \tilde{g} & \tilde{g} & \tilde{\omega}_b + 2 D & 2 D \\
        -\tilde{g} & -\tilde{g} & -2 D & -\tilde{\omega}_b - 2 D
    \end{pmatrix}\, .
\end{equation}
In \figref{fig:disp_rel_closed}(a), the upper and lower excitation energies of the standard (closed system) Dicke model are plotted as a function of the normalized coupling strength $g/\omega_a$.
Similarly, \figref{fig:disp_rel_closed}(b) presents these excitation energies as a function of the frequency ratio $\omega_b / \omega_a$, which can be more easily controlled in a typical experimental realization \cite{kono2024observation}, as it will be further discussed in the Section \ref{sec:theory_spectra}.

\section{Open Dicke Model}
\label{sec:open_dicke}

In this section, we extend the well-established results regarding the closed Dicke model and the SPT, including the interactions with the external environment. This aspect is crucial for accurately reproducing experimental results, especially in systems that are not driven-dissipative.
Indeed, while many approaches have been developed during the years for the study of the driven-dissipative Dicke model, the lack of a general theory able to describe a non-driven QPT which includes the interaction with the external environment, not relying on those approximations which fail in proximity of the critical point (\textit{e.g.}, Born-Markov and rotating wave approximations), is still missing, at least to our knowledge.

We present here a general theoretical framework based on the Quantum Langevin equations, valid both in the normal and superradiant phase, which is approximation-free, thus suitable for the description of the non-driven dissipative QPTs near their critical points. Furthermore, this approach enables the calculation of reflection, absorption, and transmission spectra for arbitrary values of the coupling strength and loss rates, in presence of arbitrary types of baths (\textit{e.g.}, ohmic and non-ohmic). 
In our derivation, we consider the general case in which both the subsystems interact with their respective external environments through the decay rates $\gamma_a$ and $\gamma_b$, respectively (see \figref{fig:ref_system}). This treatment can be easily extended in the case of additional loss channels, such as additional input-output ports or non-radiative losses, which have not been included here for the sake of simplicity. The external baths are modeled as infinite, discrete collections of independent harmonic oscillators, coupled to the system Hamiltonian through their coordinates \cite{GardinerZoller_book}. Hence, the total Hamiltonian is given by
\begin{equation} \label{eq:H_tot}
    H = H_{\rm sys} + \frac{1}{2}\displaystyle \sum\limits_{j = a, b} \sum_n \left[ p_{jn}^2 + k_{jn}(q_{jn} - X_j)^2 \right] \, ,
\end{equation}
where $q_{jn}$ and $p_{jn}$ are the coordinate and momentum associated to the $n$-th mode of the $j$-th oscillator, respectively, and $X_j$ are system coordinates. 
This form of coupling is physically well-grounded, as it admits a clear and straightforward interpretation: the bath coordinates $q_{jn}$ are shifted from their unperturbed equilibrium position by the influence of the system coordinates $X_j$.
This interaction potential has the additional advantage of displaying a metastable energy minimum. On the other hand, another commonly employed form of the system-bath interaction term is given just by a sum of products between the system and bath coordinates, \textit{i.e.}, $q_{jn} X_j$. Differently from the expression in \eqref{eq:H_tot}, the latter coupling term does not display an energy minimum and can thus introduce additional instabilities in the system description, as further discussed in Section \ref{sec:qX}, where we will show that the form of the interaction potential can have relevant consequences on the SPT of the open Dicke model.
Throughout this paper, we base our analysis of the open system dynamics on the quantum Langevin equations derived from \eqref{eq:H_tot}, without introducing any additional approximation to the system-bath interaction. An equivalent alternative approach involves the Fano-Hopfield-Bogoliubov diagonalization of the full system-bath Hamiltonian (see, e.g., Refs \cite{Savasta1996, Cortese2022,deliberato2017}).

\subsection{Normal Phase} \label{sec:normal_phase}

The coordinates of the systems $A$ and $B$ appearing in \eqref{eq:H_tot} are, respectively, $X_a = \sqrt{\hbar/2\omega_a}(a^\dagger + a)$ and $X_b = \sqrt{\hbar/2\omega_b} ( b^\dagger \sqrt{1-b^\dagger b/N} + \sqrt{1-b^\dagger b/N} \, b)$. In the thermodynamic limit, the ground state in the normal phase is not macroscopically occupied, hence, the square root in the definition of $X_b$ approaches unity, leading to the effective coordinate $X_b = \sqrt{\hbar/2\omega_b} (b^\dagger + b)$.
We can effectively trace out the bath degrees of freedom from \eqref{eq:H_tot} and derive the corresponding quantum Langevin equations (see App.~\ref{sec:quant_lang_eq_normal}) which, in the Fourier domain, are 
\begin{equation} \label{eq:Langevin_eq_input_full}
    -i \omega \tilde{\bf v} (\omega) = -i \left({\bf A} - \frac{i}{2} {\bf \Gamma}(\omega) \right) \tilde{\bf v}(\omega) + \tilde{{\bf F}}_{\rm in}(\omega) \, ,
\end{equation}
where $\tilde{\bf v} = (\tilde{a}, \tilde{a}^\dag, \tilde{b}, \tilde{b}^\dag)^T$ and $\tilde{{\bf F}}_{\rm in}$ is the Langevin forces vector of the input fields in the frequency domain. The decay matrix $\bf \Gamma$ depends on how the losses of the system are modeled, which in turn are microscopically linked to the system-bath couplings $k_{jn}$. 
By introducing the matrix ${\bf M}(\omega; {\bf A, \Gamma}) = {\bf A} - i {\bf \Gamma}(\omega)/2 - \omega {\bf I}$, \eqref{eq:Langevin_eq_input_full} can be compactly written as
\begin{equation} \label{eq:Langevin_eq_input}
    i {\bf M}(\omega; {\bf A, \Gamma}) \tilde{\bf v}(\omega) = \tilde{{\bf F}}_{\rm in}(\omega) \, .
\end{equation}
To simplify the notation, from now on we will suppress the explicit dependence on $\omega$ whenever it is not strictly necessary.
For completeness, we note that an analogous equation applies to the output fields, namely
\begin{equation} \label{eq:Langevin_eq_output}
    i {\bf M}({\bf A, -\Gamma}) \tilde{\bf v} = \tilde{{\bf F}}_{\rm out} \, .
\end{equation}
For the specific cases of only one bath for each of the two subsystems $A$ and $B$ (see \figref{fig:ref_system}), the resulting decay matrix ${\bf \Gamma}$ can be expressed in the normal phase as
\begin{equation}
    {\bf \Gamma}_{\rm NP} =
    \begin{pmatrix}
        \gamma_a & -\gamma_a & 0 & 0 \\
        -\gamma_a & \gamma_a & 0 & 0 \\
        0 & 0 & \gamma_b & -\gamma_b \\
        0 & 0 & -\gamma_b & \gamma_b
    \end{pmatrix} \, .
\end{equation}
The off-diagonal elements originate from the counter-rotating terms in the system-bath interaction in Hamiltonian in \eqref{eq:H_tot}.

If the system is not coherently driven, the mean values of the input fields vanish. Consequently, by averaging \eqref{eq:Langevin_eq_input} and imposing the compatibility constraint on the system of equations, we obtain
\begin{equation}\label{eq:zeta_func}
    \zeta (\omega; \omega_a, \omega_b, \gamma_a, \gamma_b) \equiv \det \left[ {\bf M}(\omega; {\bf A, \Gamma}) \right] = 0\, .
\end{equation}
The zeros of this equation, $\Omega$, correspond to the complex eigenfrequencies of the equilibrium open Dicke model, since $\zeta (\omega)$ is, by construction, the characteristic polynomial of ${\bf A} - i {\bf \Gamma}/2$.
In the normal phase, \eqref{eq:zeta_func} is to be evaluated using the corresponding matrices ${\bf A}_{\rm NP}$ and ${\bf \Gamma}_{\rm NP}$, which leads to $\zeta_{\rm NP} (\omega) = \det \left[ {\bf M}(\omega; {\bf A_{\rm NP}, \Gamma_{\rm NP}}) \right]$. The behavior of $\zeta_{\rm NP} (\omega)$ is displayed in \figref{fig:disp_rel} on the yellow background.
A detailed discussion on the complex eigenfrequencies will be presented in Sec.~\ref{sec:excitation_energies}.
For now, we simply observe that the critical point of the open system can be identified in a similar way to the closed Dicke model, namely through the vanishing of one of the complex eigenfrequencies. We notice that the real part of the lower complex eigenfrequency approach zero before the critical point (of the closed system), while simultaneously its imaginary part splits and reaches zero precisely at the closed-system critical value of the parameter.
Beyond this critical point, the ground state of the system acquires a macroscopic occupation, requiring a revised description of the system. We will delve into these implications in the following section.

\subsection{Superradiant Phase} \label{sec:superradiant_phase}

To accurately describe the properties of the open Dicke model in the superradiant phase, it is essential to account for the condensates that form in the system's ground state. A comprehensive and self-consistent treatment requires direct analysis of the total Hamiltonian given in \eqref{eq:H_tot}, which contains both the system and bath variables. Indeed, this approach ensures a proper description of the mutual influence between the system and its environment, which becomes relevant in the strong system-bath coupling regime.
This methodology significantly differs from previous studies, which primarily rely on either the minimization of the effective system's degrees of freedom through a master equation approach \cite{Brandes_openDicke2013, FredrikLambert2024} or the analytic continuation of the system’s Green function \cite{Roche2025}. In contrast, our approach enables a more rigorous treatment of the open-system dynamics, capturing key effects that can emerge due to the strong system-bath interactions, which becomes inevitable near an equilibrium QPT.
To perform our analysis of the superradiant phase, we proceed similarly to the standard Dicke model by shifting both the system and bath bosonic operators as
\begin{equation} \label{eq:bosonic_shifts}
\begin{aligned}
    a & = a_s + \sqrt{\alpha}\,, \quad\quad\quad\quad\;\;\;
    b = b_s - \sqrt{\beta}\,, \\
    c_{an} & = c_{{\rm s}, an} + \sqrt{\sigma_{an}}\,, \quad\quad
    c_{bn} = c_{{\rm s}, bn} - \sqrt{\sigma_{bn}} \,\, ,
\end{aligned}
\end{equation}
where, as usual, the bosonic operator of bath $A$ is defined as $c_{jn} = \left( \omega_{jn} q_{jn} + i p_{jn} \right) / \sqrt{2 \hbar \omega_{jn}}$. In principle, the baths can have different dimensionalities.
By substituting these definitions into the Hamiltonian in \eqref{eq:H_tot} and minimizing the resulting expression, we obtain the non-zero macroscopic occupations (see App.~\ref{sec:minimization_H_superradiant})
\begin{equation} \label{eq:macroscopic_occupations}
    \begin{aligned}
        \alpha & = \frac{N g^2}{\omega_a^2}\! \left( 1 - \frac{1}{\lambda^2} \right) \,,\quad\,
        \beta = \frac{N}{2} \left( 1 - \frac{1}{\lambda} \right), \\
        \sigma_{an} & =  \frac{k_{an}}{\omega_{an} \omega_a}\alpha \,,\quad\quad\quad\;\;
        \sigma_{bn} =  \frac{k_{bn}}{\omega_{bn} \omega_b}\frac{\lambda + 1}{2\lambda} \beta \,\, .
    \end{aligned}
\end{equation}
Notably, the minimization procedure reveals that the open Dicke model predicts a macroscopic ground state condensation in the superradiant phase which is exactly the same as in the isolated system, as evidenced by the same condensates' values as in the Sec.~\ref{sec:standard_Dicke}. These results highlight what can be described as a \emph{resilience} of the condensates against the effects of system-bath coupling. 
Furthermore, although the ground state condensation of the system remains unaffected by the presence of the baths, it nevertheless induces a macroscopic occupation in the bath fields. This phenomenon is to be expected in equilibrium conditions and can be interpreted as the baths inheriting a property of the system, much like a material develops magnetization when placed in contact with a magnet.
The resilience of the QPT to the interaction of the system with the external baths is due to the specific form of the system-bath coupling introduced in \eqref{eq:H_tot}. 
As discussed in detail in Sec.~\ref{sec:qX}, the form of the microscopic system-bath interaction can significantly influence the predicted outcomes. For instance, in the present case, the compensation among some of the terms involved in the minimization process, which in turn plays a crucial role in determining the macroscopic mode occupations, is directly tied to this choice.

\begin{figure}[t]
    \centering
    \includegraphics[width=\linewidth]{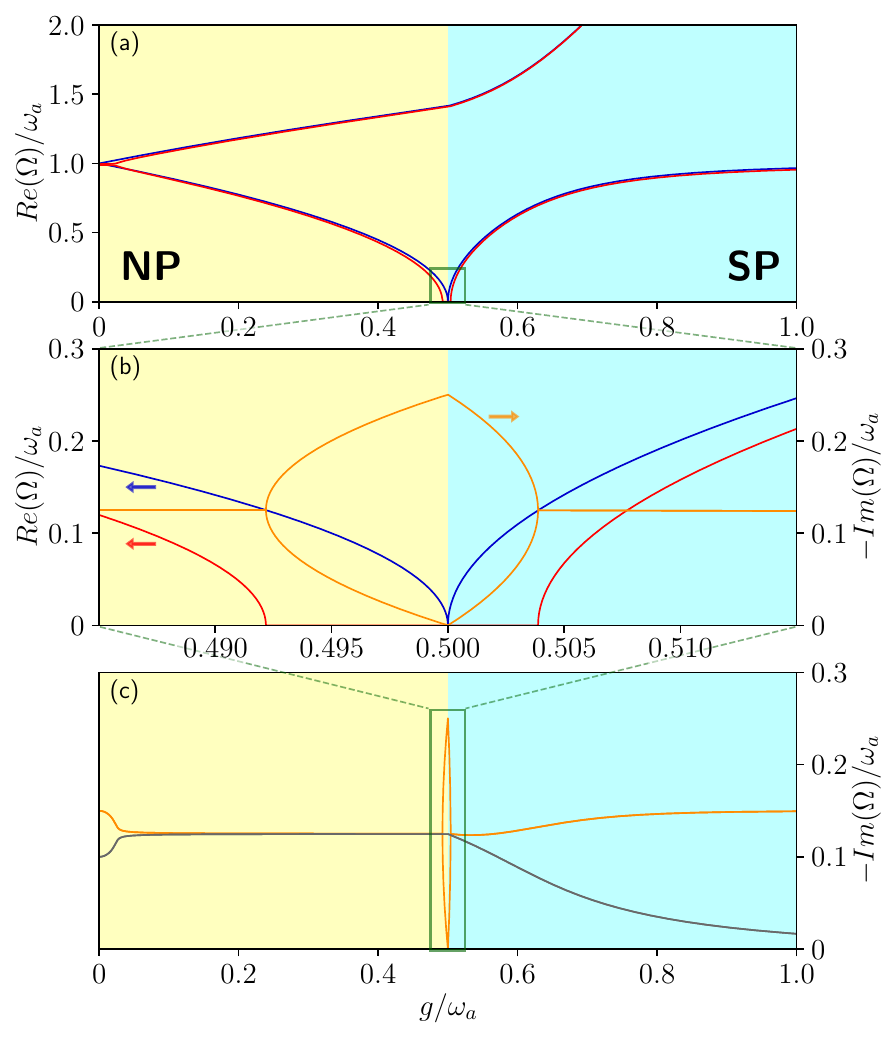}
    \caption{\textbf{Excitation energies.} \textbf{(a)} Upper and lower excitation energies as a function of the normalized coupling in the closed (blue) and open (red, real parts) Dicke models.
    \textbf{(b)} Inset zooming near the critical point of the lower polaritons of both the closed and open Dicke models.
    \textbf{(c)} Negative imaginary parts of the upper (gray) and lower (orange) excitation energies.
    Parameters used: $\omega_a = \omega_b = 1$, $\gamma_a = 0.3$, $\gamma_b = 0.2$.}
    \label{fig:disp_rel}
\end{figure}

After enforcing the equilibrium condition, the resulting effective Hamiltonian in the superradiant phase can be expressed as
\begin{equation} \label{eq:H_superradiant}
    H = H_{\rm SP} + \frac{1}{2} \sum_{j=a,b} \sum_{n} \left[p_{jn}^2 + \tilde{k}_{jn} \left( q_{{\rm s},jn} - x_j \right)^2 \right] \, ,
\end{equation}
where we defined the coordinate fluctuation operators $q_{{\rm s},jn} = \sqrt{\hbar \omega_{jn}/ 2 k_{jn}} \left( c_{{\rm s},jn}^\dagger + c_{{\rm s},jn} \right)$. 
The effective system coordinates coupled to the baths are given by $x_a =  \sqrt{\hbar/2\omega_a} (a_{\rm s}^\dagger + a_{\rm s})$ and $x_b =  \sqrt{2\hbar/\omega_b (\lambda+1)^2} (b_{\rm s}^\dagger + b_{\rm s})$, while the coupling constants are $\tilde{k}_{an} = k_{an}$ and $\tilde{k}_{bn} = k_{bn} \left( \lambda + 1 \right) / 2 \lambda$. The effects of condensation are incorporated through the $\lambda$-dependent factors.
As shown in App.~\ref{sec:quant_lang_eq_superradiant}, the quantum Langevin equations in the superradiant phase preserve the same structure as in \eqref{eq:Langevin_eq_input}, upon the introduction of the decay matrix ${\bf \Gamma}_{\rm SP}$, obtained by the replacement $\gamma_b \to \tilde{\gamma}_b = 2 \gamma_b / \lambda (\lambda + 1)$ in ${\bf \Gamma}_{\rm NP}$.
The effective loss rate of subsystem $B$, $\tilde{\gamma}_b$, coincides with the bare damping rate $\gamma_b$ at the critical point but gradually decreases to zero as the coupling strength increases. This behavior arises from saturation effects, which become increasingly significant and are determined by the values of the condensates.
Similarly to the normal phase, the excitation energies in the superradiant phase can be computed by adapting \eqref{eq:zeta_func}, which yields to $\zeta_{\rm SP} (\omega) = \det \left[ {\bf M}(\omega; {\bf A_{\rm SP}, \Gamma_{\rm SP}}) \right]$. The zeros of $\zeta_{\rm SP} (\omega)$ are plotted in \figref{fig:disp_rel} on the cyan background.

\subsection{Excitation Energies} \label{sec:excitation_energies}

Figure \ref{fig:disp_rel} displays the real (red) and the opposite of the imaginary parts (orange and gray, corresponding to the lower and upper polaritons, respectively) of the complex excitation energies obtained from \eqref{eq:zeta_func}. The most notable distinction from the excitation spectra of the closed Dicke model (blue) is the emergence of a {\em gap} region between the normal and superradiant phases (\figref{fig:disp_rel}b), where the real part of the lower polariton becomes zero while its imaginary part splits. We identify the critical point of the QPT with the vanishing of the imaginary part of a complex excitation energy, which necessarily coincides with the vanishing of the total complex eigenfrequency. It is important to highlight that extending the normal-phase excitation frequencies beyond the critical point would incorrectly predict a positive imaginary component for the lower mode, violating the principle of causality, which requires the response functions to not have any poles in the upper half of the complex frequency plane.
Indeed, in the next section, we will show that the poles of a causal response function in the frequency domain (the reflection coefficient $S_{11}$) actually coincide with the solutions of \eqref{eq:zeta_func}, thereby validating the present analysis.
Another remarkable feature is the behavior in the energy spectrum at low coupling strengths, where the imaginary parts of the upper and lower excitation energies split, while their real parts tends to converge. This characteristic behavior marks the transition between the weak and strong coupling regimes. Notably, if one of the channel effectively acts as a gain that compensates for the system's losses, \textit{i.e.}, $\gamma_b = - \gamma_a$, the effective Hamiltonian in the normal phase would display $\mathcal{PT}$ symmetry with the transition associated to an exceptional point \cite{Bender1998, Bender2007, feng2017non, Longhi2009, wennerPRL2014, sunPRL2014}.

Remarkably, it can be shown that the critical point in this equilibrium open Dicke model coincides with that of the corresponding closed Dicke model, provided the dissipation rates are {\em well-behaved}, even if not ohmic. 
To clarify this point, we analyze the behavior of the complex excitation energies in the normal phase by examining the explicit expression of $\zeta_{\rm NP}(\omega)$, given by
\begin{equation} \label{eq:zetaNP}
    \begin{aligned}
        \zeta_{\rm NP}(\omega)=&\;\omega^4 \!+i(\gamma_a + \gamma_b)\omega^3 -(\omega_a^2 + \omega_b^2 + \gamma_a\gamma_b)\omega^2 \\
    &-i(\omega_a^2 \gamma_b +\omega_b^2 \gamma_a)\omega -4g^2\omega_a\omega_b + \omega_a^2\omega_b^2 \, .
    \end{aligned}
\end{equation}

\begin{figure}[t]
    \centering
    \includegraphics[width=\linewidth]{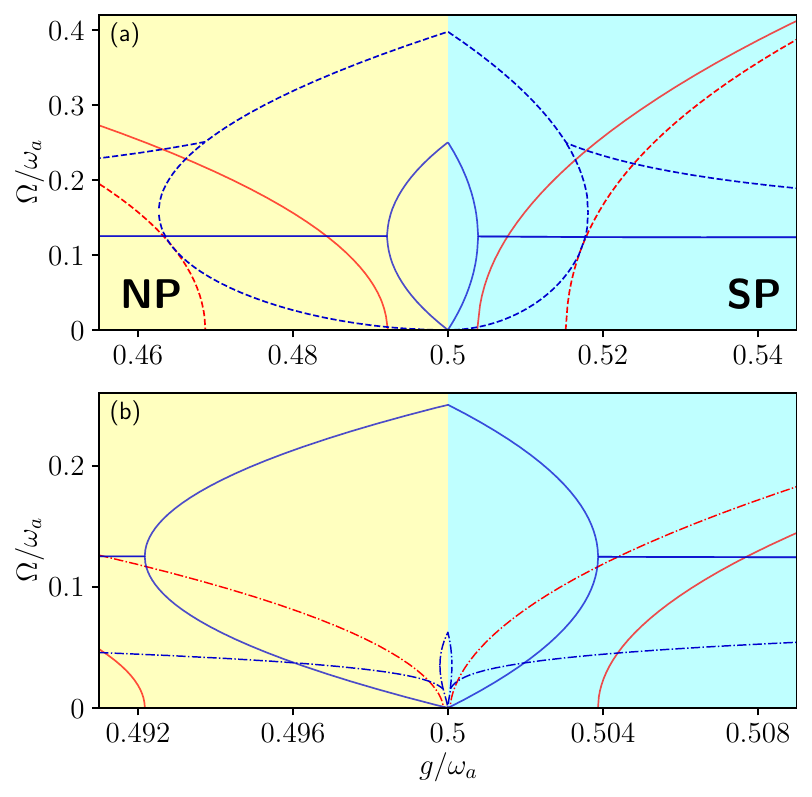}
    \caption{\textbf{Excitation energies for non-ohmic baths.} Comparison of the real (red) and imaginary (blue) parts of the lower excitation energy for ohmic baths (solid lines) with those for \textbf{(a)} subohmic baths (dotted lines) and \textbf{(b)} superohmic baths, plotted as functions of the normalized coupling strength $g/\omega_a$. Parameters used: $\omega_a = \omega_b = 1$, $\gamma_{0a} = 0.3$, $\gamma_{0b} = 0.2$.}
    \label{fig:disp_rel_sub-super}
\end{figure}

As can be readily observed, in the limit $\omega \to 0$ (\textit{i.e.}, near the QPT), only the constant terms in \eqref{eq:zetaNP} are relevant, provided the damping rates are physically meaningful. By this, we refer to the constraint that $\omega \gamma(\omega)$ must vanish as $\omega \to 0$, though the rate at which it approaches zero may vary \cite{Leggett1987}.
These constant terms vanish when the coupling strength reaches $g = \sqrt{\omega_a \omega_b}/2 = g_c$, which coincides with the critical coupling of the SPT in the closed Dicke model. This demonstrates the resilience of the SPT against the coupling with the external environments.
Although \eqref{eq:zetaNP} remains valid for any well-behaved dissipation rates, the case of constant decay rates ($\gamma_j(\omega) \equiv \gamma_{0j}$, for frequencies well below a high-frequency cutoff) is of particular interest. This corresponds to the ohmic dissipation described in Ref.~\cite{Leggett1987}, as it yields the familiar velocity-dependent damping term of a classical damped harmonic oscillator. Indeed, considering as an example the decoupled ($g=0$) oscillator $A$, the frequency-domain equation of motion reads: $-i \omega \tilde{P}_a = -\omega_a^2 \tilde{X}_a - i  \omega \gamma_a  \tilde{X}_a + \tilde \xi_a$.
Since the damping rates must satisfy the condition $\gamma_j^* (\omega) = \gamma_j(-\omega)$ (see App.~\ref{sec:quant_lang_eq_normal}), we model their low-frequency behavior as $\gamma_j(\omega) = \gamma_{0j} \abs{\omega}^s$, where $s=0$ corresponds to the ohmic case. 
Baths with $-1<s<0$, where the damping rate vanishes more slowly as $\omega \to 0$ than in the ohmic case, are classified as subohmic, while those with $s>0$ are referred to as superohmic. Pathological cases with $s\leq-1$ are excluded from the present analysis.
Remarkably, due to the form of $\gamma_j(\omega)$, the behavior of $\zeta_{\rm NP}(\omega)$ near the QPT remains unaffected, leaving the critical point unchanged even in presence of non-ohmic baths.
This result sharply contrast with previous findings for the driven-dissipative Dicke model, where the critical point is shifted by the dissipation rate \cite{Brandes_openDicke2013, FredrikLambert2024, DallaTorre2013}.
\figref{fig:disp_rel_sub-super} shows the real (red) and imaginary (blue) parts of the lower eigenfrequency for subohmic (dotted lines) and superohmic (dash-dotted lines) baths, in comparison with the ohmic case (solid lines). 
The damping rates are intentionally set to high values to emphasize the relative differences between the various types of baths. This choice is allowed by the fact that our approach does not impose any constraints on the smallness of the damping rates.

\section{Input-Output Fields and Spectra} \label{sec:theory_spectra}

\begin{figure*}[t]
    \centering
    \includegraphics[width=\textwidth]{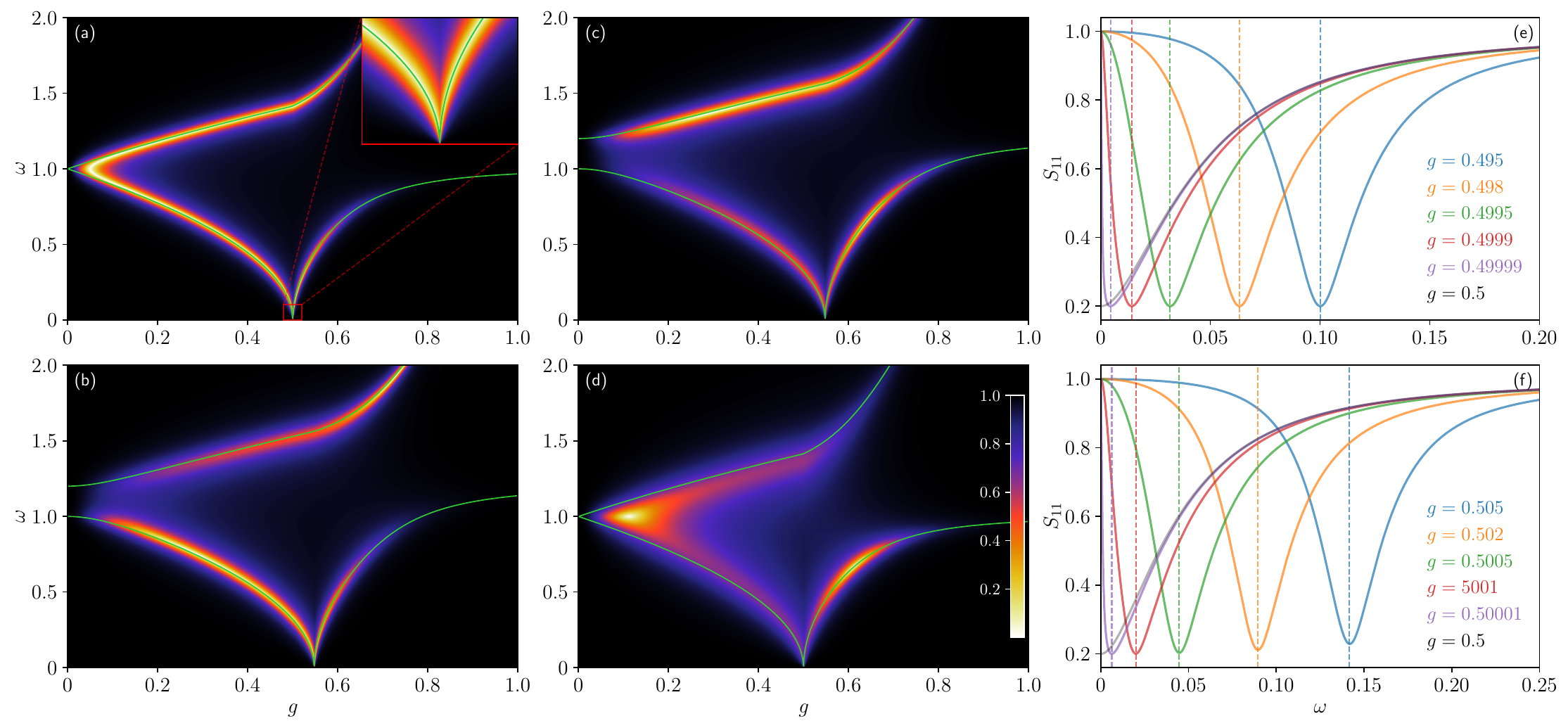}
    \caption{\textbf{Ohmic reflection spectra.} \textbf{(a-d)} 2D ohmic reflection spectra, with an inset near the critical point in (a). The corresponding closed-system excitation energies, $\Omega$, are superimposed on the plots (green lines). 
    \textbf{(e,f)} 1D spectra in the proximity of the critical point, in the (e) normal and (f) superradiant phases. Dashed vertical lines represent the corresponding closed-system lower excitation energies ($\Omega_-$), showing excellent agreement with the minima of the reflection spectra.
    Parameters used: (a) $\omega_a = \omega_b = 1$, $\gamma_a = \gamma_b = 0.1$; (b) $\omega_a =1.2$, $\omega_b = 1$, $\gamma_a = 0.2$, $\gamma_b = 0.1$; (c) $\omega_a =1.2$, $\omega_b = 1$, $\gamma_a = 0.1$, $\gamma_b = 0.2$; (d) $\omega_a = \omega_b = 1$, $\gamma_a = 0.1$, $\gamma_b = 0.5$; (e,f) $\omega_a = \omega_b = 1$, $\gamma_a = 0.05$, $\gamma_b = 0.075$.}
    \label{fig:ohmic_spectra}
\end{figure*}

The approach presented in this Section offers several advantages. Beyond its applicability across a broad range of parameters 
and the possibility of addressing non-ohmic baths, it also allows for the analytical calculation of reflection and transmission spectra for both ohmic and non-ohmic environments.
In this section, we develop the theoretical framework required for such analysis, while in the following subsections, we apply it to ohmic, subohmic, and superohmic cases, illustrating the results through examples of reflection spectra.

Following the approach outlined in Refs.~\cite{GardinerZoller_book, Yurke1984networktheory}, we introduce the input fields' operators as
\begin{equation}\label{eq:C_in}
     C_{{\rm in}, j}(t) = \!\int_0^\infty \!\!\! \sqrt{\frac{\hbar}{4 \pi \omega}} \! \left( \tilde{c}_{{\rm in}, j}(\omega) e^{-i \omega t}\! + \tilde{c}_{{\rm in}, j}^\dagger (\omega) e^{i \omega t} \right) d\omega \, ,
\end{equation}
with a similar relation holding for the output fields, $C_{{\rm out}, j}(t)$.
The creation and annihilation operators of these fields are strictly related to the baths bosonic operators, and satisfy the canonical commutation relations $\left[ \tilde{c}_{{\rm in/out}, j}(\omega^\prime) \, , \tilde{c}_{{\rm in/out}, k}^\dagger (\omega^{\prime\prime}) \right] = \delta_{jk} \delta(\omega^\prime - \omega^{\prime\prime})$.
For convenience, we adopted the continuum limit on the bath degrees of freedom. We observe that, in absence of the RWA, both co-rotating and counter-rotating terms in $C_{{\rm in}, j}(t)$ must be considered.
It can be shown that, in the normal phase, the Langevin forces are proportional to the fields in the frequency domain, \textit{i.e.}, $\tilde{F}_{{\rm in/out}, j}(\omega) \propto\sqrt{\gamma_j/\omega_j} \, \tilde{C}_{{\rm in/out}, j}(\omega)$. A similar relation holds in the superradiant phase, upon the introduction of a prefactor $\sqrt{2 / \lambda (\lambda + 1)}$ preceding $\tilde{C}_{{\rm in/out}, b}$. A detailed discussion on the derivation and properties of the input and output fields is provided in App.~\ref{sec:app_inout_op}.
The input-output relations \cite{GardinerZoller_book, WallsMilburn_book_input-output} can be obtained through Eqs.~(\ref{eq:Langevin_eq_input}) and (\ref{eq:Langevin_eq_output}), reformulated in terms of the input and output fields $\tilde{C}_{{\rm in/out}, j}$, yielding to
\begin{equation} \label{eq:output_vs_input}
    \tilde{\bf C}_{{\rm out},j} = \sum_k \sqrt{\frac{\omega_j}{\omega_k} \frac{\gamma_k}{\gamma_j}} \left. {\bf M}\left({\bf A, -\Gamma}\right) {\bf M}\left({\bf A, \Gamma}\right)^{-1} \right\vert_{jk} \tilde{\bf C}_{{\rm in},k}\,,
\end{equation}
where $\left. {\bf M}\left({\bf A, -\Gamma}\right) {\bf M}\left({\bf A, \Gamma}\right) ^{-1} \right\vert_{jk}$ is the $(j,k)$-th $2 \times 2$ block of the matrix ${\bf M\left({\bf A, -\Gamma}\right) M}\left({\bf A, \Gamma}\right)^{-1}$, for $j,k =1,2 \, (= a,b)$. The input (output) vectors are defined as $\tilde{\bf C}_{{\rm in/out}, j} = (\tilde{C}_{{\rm in/out}, j} \, , -\tilde{C}_{{\rm in/out}, j})^T$. 
Thus, we can now introduce the scattering matrix ${\bf S}(\omega)$, whose elements are defined by
\begin{equation} \label{eq:S_jk}
    S_{jk} = \left. \frac{\langle \tilde{C}_{{\rm out}, j} \rangle}{\langle \tilde{C}_{{\rm in}, k} \rangle} \right\vert_{\langle \tilde{C}_{{\rm in}, i} \rangle = 0 \,\, {\rm for}\,\, i \ne k}
\end{equation}

As already highlighted, a key advantage of this approach lies in the absence of any approximation in the derivation of \eqref{eq:S_jk}, which makes it suitable for describing the system properties across all range of the parameters, including highly nontrivial cases such as those in the proximity of the critical point or in presence of strong dissipation.

As illustrative examples, in the following sections we compute reflection spectra through port $a$ $(S_{11})$, obtained by evaluating \eqref{eq:output_vs_input} for $j=k=1$, which leads to
\begin{equation} \label{eq:S_11}
    \begin{aligned}
        S_{11} \!=&\, \left. \frac{\langle \tilde{C}_{{\rm out}, a} \rangle}{\langle \tilde{C}_{{\rm in}, a} \rangle} \right\vert_{\langle \tilde{C}_{{\rm in}, b} \rangle = 0} \\
        =&\, \left( \!2^{-\frac{1}{2}},-2^{-\frac{1}{2}}\! \right) \! \left. {\bf M}\left({\bf A,\! -\Gamma}\right) {\bf M}\left({\bf A, \Gamma}\right) ^{-1} \right\vert_{11} \!\!
        \begin{pmatrix}
            2^{-\frac{1}{2}} \\
            -2^{-\frac{1}{2}}
        \end{pmatrix} \\
        =&\, \frac{\zeta (\omega_a, \omega_b, -\gamma_a, \gamma_b)}{\zeta (\omega_a, \omega_b, \gamma_a, \gamma_b)}\, .
    \end{aligned}
\end{equation}
As previously mentioned, the denominator in this expression corresponds to the characteristic polynomial of the open Dicke model.
Furthermore, \eqref{eq:S_11} is valid for both ohmic and non-ohmic baths.
Spectra associated with alternative input-output channels, such as the transmission spectrum $S_{12}$, can be readily obtained by appropriately applying \eqref{eq:S_jk}.

\subsection{Ohmic Reflection Spectra} \label{sec:ohmic_spectra}

In this section, we focus our attention on the analysis of ohmic spectra $(s=0)$, which corresponds to assuming constant damping rates, \textit{i.e.}, $\gamma_j(\omega) = \gamma_{0j}$ in \eqref{eq:S_11}. 
While maintaining constant damping rates over a broad spectral range is not entirely realistic, here our main interest lies in the low-frequency region, where the softening of the lower polariton occurs near the critical point. Nevertheless, the theoretical framework employed here allows for the incorporation of more complex frequency-dependent damping rates without any issues.

\begin{figure}[b]
    \centering
    \includegraphics[width=\linewidth]{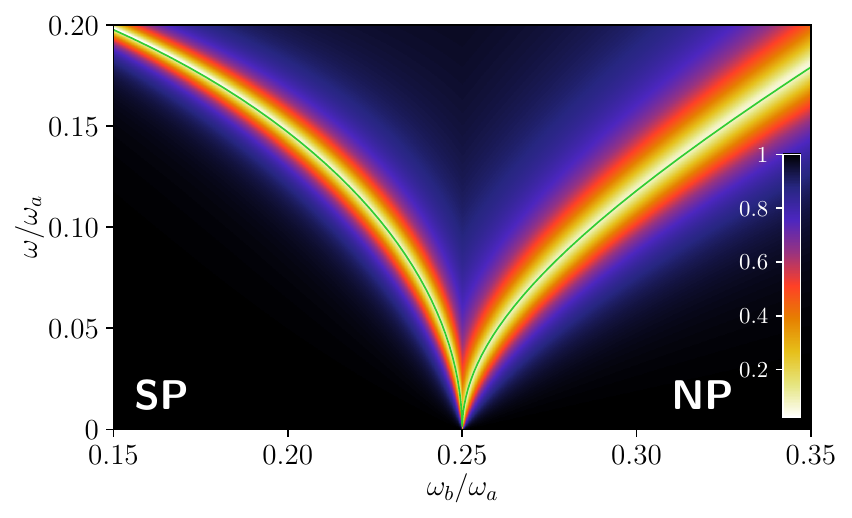}
    \caption{\textbf{Typical experimental reflection spectra.} 2D ohmic reflection spectra as a function of the frequency ratio $\omega_b / \omega_a$, for $g/\omega_a = 0.25$.}
    \label{fig:ohmic_spectra_nu}
\end{figure}

In \figref{fig:ohmic_spectra}, we plot the reflection spectra for an ohmic bath over a wide range of parameters, taking advantage of the opportunities our treatment allows for. A key feature observed is the presence of a coupling-dependent Lamb shift, which displaces the reflection minima relative to the eigenfrequencies of the isolated system (green solid lines). However, while the Lamb shift is appreciable far from the critical point, it vanishes in the vicinity of $g_c$, as highlighted in the inset of panel (a) and panels (e-f). Simultaneously, the left-side broadening of the asymmetric Lorentzian profile shrinks to zero at the same rate as the minima, ensuring a consistent spectral description even in the low-frequency regime. 
It is worth noting that the reflection spectrum exactly at the critical point is not entirely physically meaningful for the lower polariton, as it exhibits a reflectivity different from $1$ at $\omega=0$. However, this does not pose a practical issue in experiments and applications, since the critical point represents a singular set of zero measure in the coupling parameter space and cannot be precisely accessed due to various intrinsic system noises. Indeed, for any coupling $g = g_c \pm \epsilon$, where $\epsilon$ is a arbitrary small positive quantity, the spectra remain well-behaved. 
While panel (a) represents a resonant system ($\omega_a = \omega_b = 1$), a comparison of panels (b) and (c) provides additional insight: both depict detuned systems ($\omega_a = 1.2, \omega_b = 1$), but in panel (b) the subsystem A has an higher damping rate compared to $B$ ($\gamma_a = 0.2, \gamma_b = 0.1$), whereas panel (c) illustrates the opposite scenario. The most striking consequence is the difference in the depth of the reflection minima for the upper and lower excitation energies, which is to be expected due to the different nature of the polaritons at low and high coupling strengths.
Panel (d) showcases a more extreme scenario where $\gamma_b = 0.5$, placing it well outside the parameter range where the Born-Markov approximation remains valid, regardless of the coupling strength. In this case, we observe a considerable Lamb shift, particularly for the upper excitation energy and for weak couplings in the lower branch. However, even in this extreme situation, the Lamb shift vanishes when approaching the critical point for the lower polariton.
Finally, panels (e,f) present 1D reflection spectra near the critical point (solid lines), alongside the excitation energies of the closed Dicke model described by equation \eqref{eq:Dicke_Ham_boson} (vertical dashed lines). 
Remarkably, these plots confirm that the reflection minima near the critical point do not correspond to the real part of the complex eigenfrequencies, but are determined by the closed-system excitation energies. This feature explains why, even near a QPT, experimental fits remain accurate when compared with the conventional closed-system eigenfrequencies (see, e.g., \cite{kono2024observation}). Furthermore, this analysis aligns with the previous observation of the vanishing of the Lamb shift near the critical point, both from below (e) and above (f).

\begin{figure*}[t]
    \centering
    \includegraphics[width=0.85\textwidth]{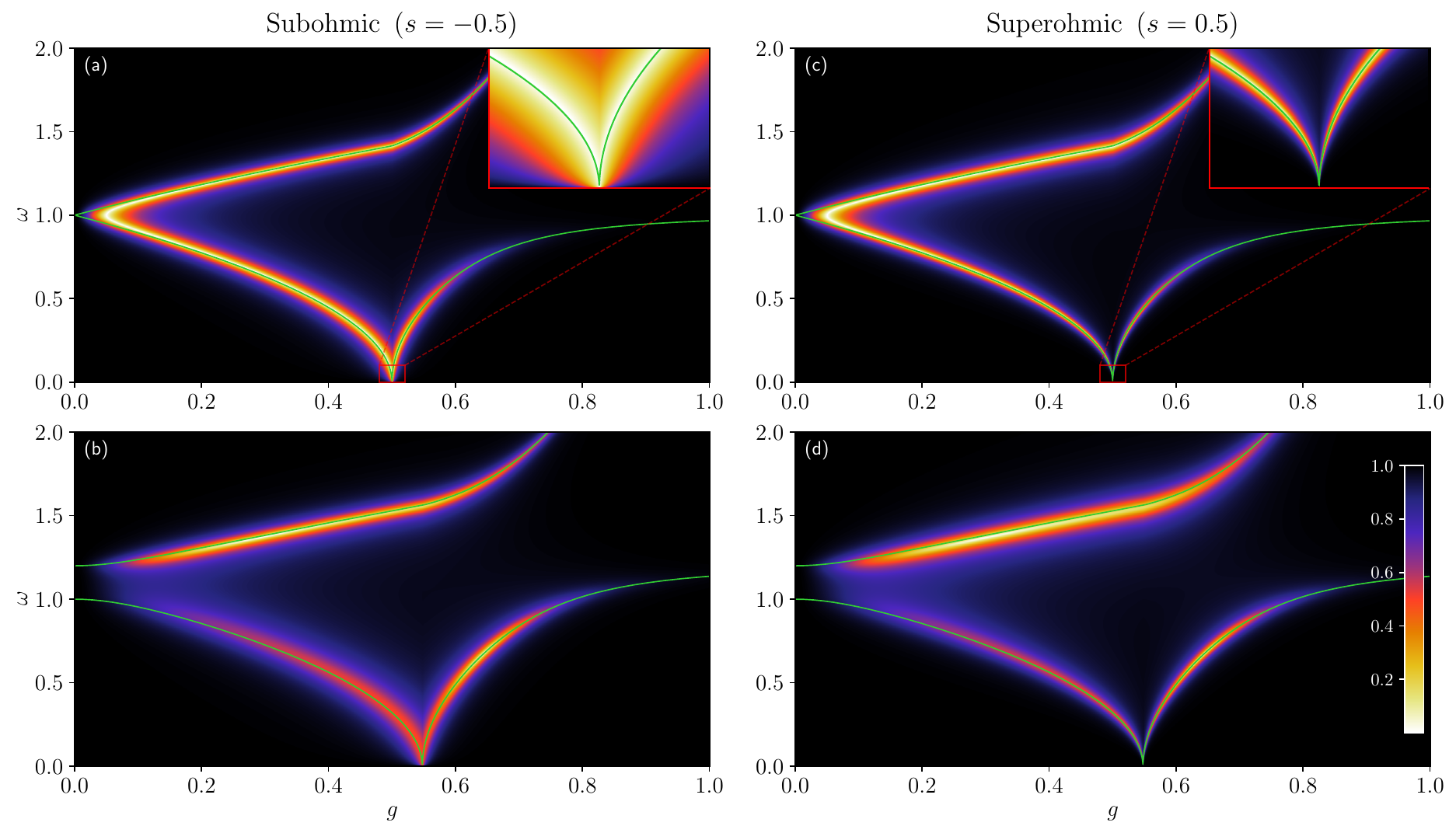}
    \caption{\textbf{Non-ohmic reflection spectra.} (a-d) 2D reflection spectra for subohmic (a,b) and superohmic (c,d) baths, with the corresponding closed-system excitation energies, $\Omega$ (green lines). 
    The insets near the critical point highlight the differences in the spectral behavior.
    Parameters used: (a,c) $\omega_a = \omega_b = 1$, $\gamma_{0a} = \gamma_{0b} = 0.1$; (b,d) $\omega_a =1.2$, $\omega_b = 1$, $\gamma_{0a} = 0.1$, $\gamma_{0b} = 0.2$.}
    \label{fig:non_ohmic_spectra}
\end{figure*}

Experimental realizations \cite{kono2024observation}, however, are typically realized by tuning the resonance frequency of one of the two subsystems, as direct modulation of the coupling strength mid-experiment is often highly challenging. 
In particular, Ref.~\cite{kono2024observation} reports the experimental observation of a magnonic SPT in a ErFeO$_3$ system, where the harmonic and anharmonic subsystems correspond to a Fe$^3+$ magnonic mode and an Er$^3+$ spin ensemble, respectively. 
This system can be effectively described by a generalized Dicke model, and its spectral properties have been investigated by varying an externally applied magnetic field, which primarily affects the resonance frequency of the Er$^3+$ spins, $\omega_b$.
\figref{fig:ohmic_spectra_nu} presents the corresponding theoretical emission spectrum as a function of the frequencies ratio $\omega_b/\omega_a$, exhibiting a close agreement with the experimentally observed SPT.
The damping rate $\gamma_b$ is assumed to scale linearly with $\omega_b$, a reasonable assumption for an ohmic spectral density. 
We point out that this calculation has been carried out at zero temperature and does not consider any additional features beyond the standard Dicke model. Hence, it is not intended to provide a quantitative fit of the experimental data but rather to offer a theoretical perspective on the underlying spectral features.

\subsection{Non-Ohmic Spectra} \label{sec:non_ohmic_spectra}

In this section, we examine both subohmic and superohmic spectral behaviors. As illustrative examples, we select $s=-0.5$ for the subohmic case and $s=0.5$ for the superohmic case, and calculate the reflection spectra using \eqref{eq:S_11}. Spectra for other types of baths can be straightforwardly computed using the same procedure.

\begin{figure}[!b]
    \vspace*{-0.9cm}
    \centering
    \includegraphics[width=0.45\textwidth]{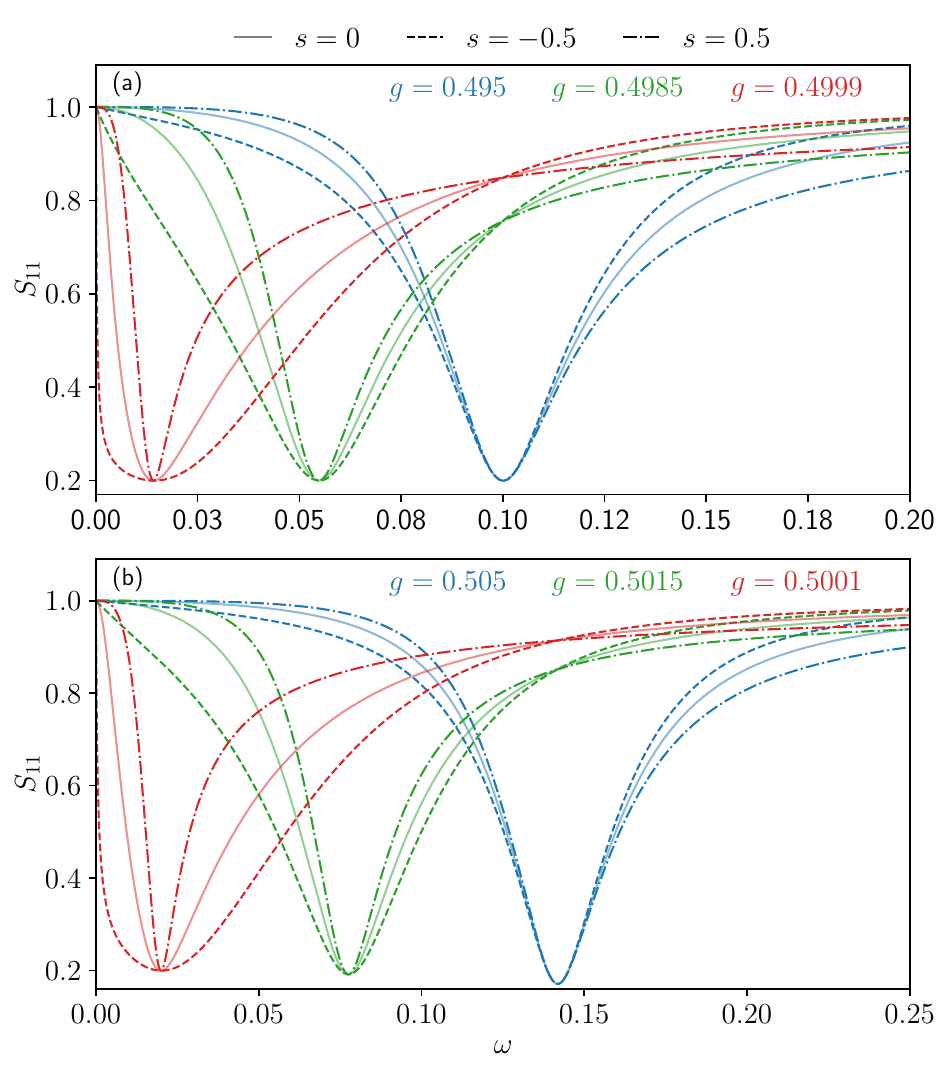}
    \caption{\textbf{Non-ohmic reflection spectra.} (a,b) Comparison of the 1D spectra in the proximity of the critical point, both in (a) normal and (b) superradiant phases, for different baths' spectral densities: ohmic (solid), subohmic (dashed) and superohmic (dash-dotted). These different behaviors have been modeled through $\omega$-dependent damping rates, $\gamma_j(\omega) = \gamma_{0j} \abs{\omega}^s$. The constants $\gamma_{0j}$ have been chosen such that the damping rates of the lower excitation energies $\gamma_j (\Omega_-)$ for the blue lines would be equal, allowing for a better comparison between the lineshapes. Parameters used: $\omega_a = \omega_b = 1$, $\gamma_{0a} = 0.05$, $\gamma_{0b} = 0.075$ for the ohmic case, while in the subohmic and superohmic cases these have been respectively divided or multiplied by (a) $3.162$ in the normal phase and (b) $2.656$ in the superradiant phase.}
    \label{fig:non_ohmic_spectra_vertical_cuts}
\end{figure}

In \figref{fig:non_ohmic_spectra}(a,b), we present subohmic reflection spectra, utilizing the same parameters as those used in panels (a) and (c) of \figref{fig:ohmic_spectra}, respectively. Similarly, \figref{fig:non_ohmic_spectra}(c,d) displays superohmic reflection spectra for the corresponding parameter set.
In the subohmic spectra, we observe a broadening of the full width at half maximum (FWHM) for the lower polariton and a narrowing for the upper polariton compared to the ohmic case. This behavior arises from the frequency-dependent scaling of the damping, $\gamma_j(\omega) = \gamma_{0j} / \sqrt{\abs{\omega}}$, which amplifies losses at low frequencies while reducing them at high frequencies.
Conversely, in the superohmic case, characterized by $\gamma_j(\omega) = \gamma_{0j} \sqrt{\abs{\omega}}$, losses are enhanced at high frequencies and suppressed at low frequencies.
This behavior is particularly evident in the insets near the critical point shown in \figref{fig:non_ohmic_spectra}(a,c), especially when compared to the inset of \figref{fig:ohmic_spectra}(a) for the ohmic bath. The range of all three insets is the same to better highlight the differences in spectral characteristics.
\figuref{fig:non_ohmic_spectra_vertical_cuts}(a,b) display 1D lower polariton spectra for coupling strengths close to the critical point, showing the differences between various baths both (a) below and (b) above the QPT.
The most notable difference lies in the distinct lineshapes between different baths, which cannot be solely attributed to variations in FWHM. Instead, it directly arises from the frequency dependence of $\gamma_j (\omega)$, leading to a highly asymmetric Lorentzian behavior.
To better emphasize this feature, the damping rates in \figuref{fig:non_ohmic_spectra_vertical_cuts}(a,b) have been deliberately chosen to differ across the illustrated examples. Specifically, the values of $\gamma_{0j}$ for the different baths have been adjusted so that the damping rates evaluated at the lower eigenfrequency, \textit{i.e.}, $\gamma_j (\Omega_-)$, are all equal for  specific coupling strengths near the critical point ($g = 0.495$ and $g = 0.505$ for the normal and superradiant phases, respectively, corresponding to the blue lines). This ensures a fair comparison of the spectral asymmetry originating from the frequency dependence of the damping rates.

\section{System-Bath interaction without a metastable minimum}
\label{sec:qX}

Typically, the coupling between the system of interest and the bath is not given much specific attention, with one of the most common forms being that given in \eqref{eq:H_tot}. However, in this section, we explore the consequences of adopting alternative interaction terms that do not exhibit an energy minimum. As noted in Sec.~\ref{sec:open_dicke}, interaction terms that are definite positive ensure that the bath does not introduce additional instabilities in the system’s description. Nonetheless, it is interesting to investigate other notable and widely adopted cases, such as interactions proportional to the product of system and bath coordinates, and how these alternatives influence the system properties. In the case of the bilinear interaction term in the coordinates, which is characteristic of certain circuital configurations, the total system-bath Hamiltonian takes the form
\begin{equation} \label{eq:H_tot_qX}
    H^{\prime} = H_{\rm sys} +\displaystyle \sum\limits_{j = a, b} \sum_n \left[ \frac{p_{jn}^2}{2} + k_{jn}\frac{ q^2_{jn}}{2}  - k_{jn} q_{jn} X_j \right] \, .
\end{equation}
Through straightforward algebraic manipulations, it becomes evident that \eqref{eq:H_tot_qX} can be rewritten in the form of \eqref{eq:H_tot} by appropriately redefining the system Hamiltonian $H_{\rm sys}$. Specifically, we obtain
\begin{equation} \label{eq:H_tot'}
     H^{\prime} = H^{\prime}_{\rm sys} + \frac{1}{2}\displaystyle \sum\limits_{j = a, b} \sum_n \left[ p_{jn}^2 + k_{jn}(q_{jn} - X_j)^2 \right] \, ,
\end{equation}
where the renormalized system Hamiltonian is given by
\begin{equation} \label{eq:H'}
    H^{\prime}_{\rm sys}=H_{\rm sys} - \sum_{j=a,b} \frac{f_j(0)}{2} X_j^2 \, .
\end{equation}
Here, $f_{j}(0)=\sum_n k_{jn}$, as defined in App.~\ref{sec:quant_lang_eq_normal}.

As emphasized in Sec.~\ref{sec:open_dicke}, a system-bath coupling of the form in \eqref{eq:H_tot'} does not alter the position of the critical point compared to the closed system. Therefore, determining the location of the SPT simply requires diagonalizing \( H^{\prime}_{\rm sys} \). 
By applying a Bogoliubov transformation to the system operators (see, e.g., Ref.~\cite{Lamberto2024}), one finds that \(H^{\prime}_{\rm sys}\) is formally equivalent to \(H_{\rm sys}\) under the redefinitions of the bare frequencies \(\omega_j^{\prime \, 2} = \omega_j^2 - f_j(0)\) and of the coupling constant \(g^{\prime} = g \sqrt{\omega_a \omega_b/{\omega^{\prime}_a \omega^{\prime}_b}} \,.\)

Consequently, the critical coupling at which the phase transition occurs is given by
\begin{equation}
    g^{\prime}_c = g_c \, \sqrt[4]{\left(1-\frac{f_a(0)}{\omega_a^2}\right) \left(1-\frac{f_b(0)}{\omega_b^2}\right)} \, .
\end{equation}
The presented equation shows that, in regard to this specific system-bath interaction, the critical point for the SPT is shifted, occurring at a lower coupling than that observed in the closed Dicke model (see Sec.~\ref{sec:standard_Dicke}). This result confirms that the form of the coupling between the subsystem and the bath can exert a significant influence on the properties of the overall system under consideration.
As a direct consequence of the redefinition of the bare frequencies $\omega_j^\prime$ and of the coupling strength $g^\prime$, a further crucial modification is the alteration in the condensates as \eqref{eq:macroscopic_occupations} now involves the primed quantities.
We also observe that the interaction of a system component with the external environment, in this case, can in principle induce a QPT with an abnormal phase for $f_j(0) > \omega_j^2$. These QPTs are totally uncorrelated from the SPT and can occur independently from the strength of the coupling $g$ between the subsystems.
Hence, in contrast with system-bath interactions with a metastable minimum, in this case the interaction also affects the structural properties of the system.

\section{On the observability of intrinsic squeezing in the open Dicke model}\label{sec:squeezing}

Fundamental concepts as squeezing and entanglement have been extensively studied in quantum optics for several decades. Their importance has grown even further with advancement in various fields, including quantum computing and quantum cryptography, as well as in many-body systems displaying QPTs near their criticality \cite{Hertz1976quantumcritical,Osterloh2002,Vidal2003entanglementcritical}.
It has been demonstrated that the ground state of the Dicke model exhibits two-mode squeezing, which becomes perfect at the critical point of the QPT \cite{Brandes03, hayashida2023perfect}. This result could have significant implications for sensing applications \cite{hotter2024}, as intrinsic squeezing has the potential to enhance the robustness of quantum sensing and information technologies against photon loss and decoherence \cite{DegenQsensing,BraunsteinQinformation}. Typically, squeezing is observed with a signal of appreciable intensity, whereas the ground state does not emit. However, recent advancements in nonlinear optical techniques, particularly electro-optic sampling, have enabled the direct probing of vacuum fluctuations \cite{Riek2015direct, Benea2019electric, Lindel}. 
Since the detection of a signal requires considering an open system, the investigation of the open Dicke model becomes essential.
Our aim is to study whether this form of ground state squeezing can be experimentally detected using these recently developed techniques.

In electro-optic measurements, the output field results from the convolution of the input field with the field inside the system, each oscillating at its respective eigenfrequency. This principle is strictly correlated to energy conservation \cite{Moskalenko}. Consequently, when defining the field quadratures, it is essential to express them in terms of physical operators oscillating at the system's actual frequencies, rather than using standard bare operators. 
More specifically, bare operators ($a,b$), due to counter-rotating terms in the interaction, do not possess a well-defined frequency. Indeed, both the annihilation and creation operators contain contributions from both positive and negative frequencies \cite{Ridolfo2012, Stassi_2016}. Therefore, the analysis of electro-optic detection techniques requires defining quadratures by decomposing the fields into their positive- and negative-frequency components, denoted $X^+ (\omega)$ and $X^- (\omega)$, respectively.

As an illustrative example, let us examine the case in which the measurement apparatus is placed outside the reference system, which can be assumed to be a single-mode electromagnetic resonator in interaction with a magnetic material. 
Probing vacuum fluctuations implies a free vacuum field as input state and analyzing how its interaction with the system modifies the state.
In light of the aforementioned considerations, we define the positive- and negative-frequency operators as
\begin{equation} \label{eq:X+_X-}
    \begin{aligned}
        X^+(\omega)&= \tilde{C}_{{\rm out}, a}(\omega) \, , \\
        X^-(\omega)&= \tilde{C}^\dagger_{{\rm out}, a} (\omega) \, .
    \end{aligned}
\end{equation}
The corresponding $\phi$-dependent quadrature operator is \(X_{\phi}(\omega)=e^{i\phi}X^+(\omega)+e^{-i\phi}X^-(\omega)\), whose ground-state variance is $(\Delta X_{\phi})^2 \equiv \bra{0}(X_{\phi})^2\ket{0} - \bra{0}X_{\phi}\ket{0}^2 = \bra{0}(X_{\phi})^2\ket{0}$.

To highlight the fundamental physics without mathematical complications, in this section we analyze the problem within the dispersive regime \cite{Zueco2009dispersive}. The examination of the general case is postponed to App.~\ref{sec:Two mode sqeezing}.
In the dispersive limit, $\omega_b \gg \omega_a$, the spin dynamics can be traced out through the use of a Schrieffer-Wolff transformation \cite{Schrieffer1966Relation,Bravyi2011schriefferwolff}, resulting in an effective Hamiltonian in the normal phase
\begin{align} 
    H_{\rm SW} & =  \hbar \omega_a a^\dag a -  \frac{\hbar g^2}{\omega_b} \left( a^\dag + a \right)^2 \nonumber \\
    & + \frac{1}{2}\displaystyle \sum_n \left[ p_{an}^2 + k_{an}(q_{an} - X_a)^2 \right] \, ,
\end{align}
where the coupling operator \(X_a\) is the same as in \eqref{eq:H_tot}. Notably, the application of the Schrieffer-Wolff transformation does not alter the coupling between the system and the thermal bath, $k_{an}$.
Taking the dispersive limit of \eqref{eq:output_vs_input}, the output field reads
\begin{equation}\label{eq:outfuori}
 \tilde{C}_{{\rm out}, a}(\omega)\!=\frac{\omega \, \omega_b(\omega-i\gamma_a)\!+\omega_a( 4 g^2 \!-\omega_a\omega_b)}{\omega \, \omega_b(\omega+i\gamma_a)\!+\omega_a( 4 g^2 \!-\omega_a\omega_b)} \,\tilde{C}_{{\rm in}, a}\,(\omega) \, .
\end{equation}
In turn, the input field can be expressed in terms of the bath operators as
\begin{equation}\label{eq:inbath}
    \tilde{C}_{{\rm in}, a}(\omega) = i\sqrt{\frac{\hbar}{2 \omega}}\tilde{c}_{{\rm in}, a}(\omega) \, ,
\end{equation}
for $\omega > 0$, whereas it is proportional to the creation operator $\tilde{c}_{{\rm in}, a}^\dagger (\omega)$ for $\omega < 0$ (see App~\ref{sec:app_inout_op}).
Inserting Eqs.~(\ref{eq:outfuori}) and (\ref{eq:inbath}) into \eqref{eq:X+_X-} and recalling that \(\tilde{c}_{{\rm in}, a}(\omega)\ket{0}=0\), we obtain
\begin{equation} \label{eq:variance}
      (\Delta X_{\phi})^2 = \bra{0} X^+(\omega)X^-(\omega)\ket{0} =\frac{\hbar}{2\omega}\,.
\end{equation}
The resulting variance \((\Delta X_{\phi})^2\) is independent of the angle \(\phi\), thus demonstrating that the squeezing in the ground state cannot be detected using these techniques. 
Equation (\ref{eq:variance}) is strictly related to the experimental scheme employed in electro-optical detection.
The impossibility of measuring squeezing aligns with the findings of Ref.~\cite{Stassi_2016}. However, the analysis of that work is focused on systems where counter-rotating terms were present only within the system Hamiltonian, while the interaction with the bath was approximated using the RWA. Additionally, the detection scheme there considered was based on homodyne detection rather than electro-optic sampling.

Within the present framework, it is also possible to consider the placement of the electro-optic crystal inside the cavity. In this scenario, following an analogous derivation to that of \eqref{eq:outfuori}, it can be shown that vacuum squeezing remains undetectable. In this case, we would only observe a modification in the photonic density of states, similarly to what found in Ref.~\cite{Lindel}.

On the contrary, we expect that strong two-mode squeezing and quantum entanglement could be observed near the critical point by implementing a non-adiabatic time-modulation of the coupling strength $g$, or of one of the two bare frequencies $\omega_{a(b)}$.
This physical process reminds the dynamical Casimir effect observed in superconducting circuits \cite{Johansson2009, Wilson2011} and quantum vacuum radiation in polariton systems \cite{Liberato2007}, in the absence of QPTs. Instead, the theoretical framework presented here also provides the means to explore the effects of non-adiabatic time-modulation in proximity of a critical point.

As previously mentioned, the ground state of the Dicke model has been shown to exhibit squeezing. Here, we emphasize that analyzing this intrinsic squeezing, following the approach outlined in \cite{Brandes03,hayashida2023perfect}, is indeed possible within an open quantum system. However, it should be noted that the results will differ from those of the closed system due to an additional source of squeezing arising from the presence of counter-rotating terms in the coupling between the subsystems and their thermal baths. This form of virtual squeezing also emerges in the fundamental case of a simple harmonic oscillator interacting with its reservoir, provided no weak-coupling or RWA assumptions are applied.

\section{Conclusions} \label{sec:conclusions}

We have studied the time-independent open Dicke model in the absence of any approximation on the system-bath interactions.
The total system-bath Hamiltonian includes the Dicke Hamiltonian in the thermodynamic limit ($N\to \infty$), the  Hamiltonians  for sets of infinite number of harmonic oscillators (the bath normal modes), and the interaction term between the system components and their baths, described by harmonic potentials with (or even without, Sec.~\ref{sec:qX}) metastable minima. In particular, we focused our attention in the proximity of the critical point, where usual approximations (e.g., Born-Markov) fails.
We have investigated the superradiant phase of the open Dicke model by diagonalizing the total Hamiltonian of the open quantum system, thus ensuring a self-consistent treatment that also accounts for the baths degrees of freedom.
The system dynamics, both in the normal and superradiant phase, has been analyzed by means of proper quantum Langevin equations.


Our results show that the Dicke QPT is resilient to interactions with the environment, at least for a large class of thermal baths with a {\em well-behaved} density of states (properly vanishing as $\omega \to 0$), and displaying system-bath interactions with metastable minima. In this case, despite the presence of large Lamb shifts and significant changes of the eigenvalues, we find that the critical point is not affected by the environment, in agreement with recent experimental results \cite{kono2024observation}. Moreover, in the superradiant phase, the system ground state condensates are not affected by the external environment. In contrast, we show that the baths get infected by the system properties and acquire a ground state macroscopic occupation themselves, which can be in principle detected. This phenomenon is analogous to the magnetization induced in a material brought into interaction with a ferromagnet.

This theoretical framework allowed us to easily calculate the spectral properties of the system when probed by a coherent tone at any value of the critical parameters, for different bath spectral densities and for arbitrary large damping rates. The analytical spectra obtained display lineshapes that become increasingly asymmetric when approaching the critical point. We also find that in the superradiant phase, the effective damping rate of the anharmonic system gradually decreases as the coupling strength increases. This result can be observed in the system spectral behavior and is induced by saturation effects determined by the condensates.

This approach can be readily extended to account for additional dissipation channels, both radiative and non-radiative, or to consider nonzero temperature processes. 
Moreover, the framework here presented can be rather directly generalized to study the influence of the system-baths interactions in arbitrary materials collectively coupled to an electromagnetic resonator in all regimes of light-matter coupling, including symmetry-broken phases \cite{Roche2025}.
Notable examples could be the Lipskin-Meshkov-Glick and Fermi-Hubbard models or the cavity-modified quantum Hall effect.
Our approach could, in principle, also be applied to driven-dissipative QPTs, albeit being an overcomplication, since approximations such as the Born-Markov and rotating-wave can be safely employed in those cases. Nevertheless, we expect that this framework may still offer deeper physical insight into the microscopic nature of the system-bath interactions.
This theoretical framework also allows for the exploration of the effects of non-adiabatic time-modulation in proximity of a critical point, which is expected to lead to the detection of vacuum emission with highly nonclasiscal features.
Future studies will extend this framework to nonlinear quantum systems, including the Dicke model with a limited number of emitters. This would be relevant in the study of topological systems and quantum computations in systems predicting a QPT.
We also believe that the rigorous description of the open Dicke model in the proximity of the critical point here presented can be valuable for criticality-enhanced quantum sensing.


\section{Acknowledgments}

S.S. acknowledges financial support by the Army Research Office (ARO) through Grant No. W911NF1910065 and the National Recovery and Resilience Plan (PNRR), Mission 4, Component 2, Investment 1.4, Call for tender No. 1031 published on 17/06/2022 by the Italian Ministry of University and Research (MUR), funded by the European Union - NextGenerationEU, Project Title “National Centre for HPC, Big Data and Quantum Computing (HPC)” - Code National Center CN00000013 - CUP D43C22001240001.

\appendix

\section{Quantum Langevin Equations in the Normal Phase} \label{sec:quant_lang_eq_normal}


The aim of this Appendix is to derive quantum Langevin equations for the Dicke Hamiltonian interacting with the external environment through an interaction term as in \eqref{eq:H_tot}. In particular, here we focus on establishing the general theoretical approach and then apply it to the effective Hamiltonian in the normal phase, while the analysis of the superradiant phase will be presented in App.~\ref{sec:quant_lang_eq_superradiant}. We assume, as usual, that the baths (near their equilibrium, at least) can be described be an infinite set of independent harmonic oscillators. This model becomes exact, e.g., if we are considering electromagnetic baths.

By solving the equations of motion for the bath degrees of freedom in terms of the system variables and subsequently inserting the results in the system's Heisenberg equations, we obtain the quantum Langevin equations \cite{GardinerZoller_book}
\begin{align}\label{eqlengev}
    \dot{Y}(t) &= \frac{i}{\hbar}\left[H_{sys},Y(t)\right] \nonumber \\
    &-\! \sum\limits_{j = a, b}\frac{i}{2\hbar} \! \left[\left[X_j,Y\right], \xi_j(t) - \! \int_{t_0}^t \! \dot{X}_j(t') f_j(t-t') \, dt' \right. \nonumber \\
    & \,\qquad\qquad\qquad\qquad\left. - f_j(t-t_0)X_j(t_0) \vphantom{\int} \right]_+ \, ,
\end{align}
where \(Y\) is an arbitrary system operator, $\left[ \dots \, , \, \dots \right]_+$ is the anticommutator, $f_j(t)=\sum_n k_{jn} {\rm{cos}}(\omega_{jn}t)$ plays the role of a memory function and
\begin{equation} \label{eq:xi_ab}
    \xi_{j}(t) \!=\! \sum_n \! \sqrt{\frac{\hbar k_{jn} \omega_{jn}}{2}} \! \left( c_{jn}^\dagger(t_0) e^{i\omega_{jn}(t-t_0)} \!+ h.c. \right)
\end{equation}
are the baths noise terms.
As already pointed out in Sec.~\ref{sec:normal_phase}, the system coordinates in the normal phase are $X_a = \sqrt{\hbar/2\omega_a}(a^\dagger + a)$ and $X_b = \sqrt{\hbar/2\omega_b}(b^\dagger + b)$. Therefore, we can calculate the Langevin equations for the system's bosonic operators in the normal phase, which read
\begin{equation} \label{eq:lang_time}
    \begin{cases}
    \begin{aligned}
        \dot{a} & = -i\omega_a a -ig(b^\dagger+b) \\
        & -\frac{i}{2\omega_a}\int_{t_0}^t \! f_a(t-t')[\dot{a}^\dagger(t')+\dot{a}(t')] dt' +\frac{i}{\sqrt{2\hbar\omega_a}}\xi_a \\
        \rule{0pt}{3ex}
        \dot{a}^\dagger & = i\omega_a a^\dagger+ig(b^\dagger+b) \\
        & +\frac{i}{2\omega_a}\! \int_{t_0}^t \! f_a(t-t')[\dot{a}^\dagger(t')+ \dot{a}(t')] dt' -\frac{i}{\sqrt{2\hbar\omega_a}}\xi_a \\
        \rule{0pt}{3ex}
        \dot{b} & = -i\omega_b b -ig(a^\dagger+a) \\
        & -\frac{i}{2\omega_b}\int_{t_0}^t \! f_b(t-t')[\dot{b}^\dagger(t')+ \dot{b}(t')] dt' +\frac{i}{\sqrt{2\hbar\omega_b}}\xi_b \\
        \rule{0pt}{3ex}
        \dot{b}^\dagger & = i\omega_b b^\dagger +ig(a^\dagger+a) \\
        & +\frac{i}{2\omega_b}\int_{t_0}^t \! f_b(t-t')[\dot{b}^\dagger(t')+ \dot{b}(t')]dt' -\frac{i}{\sqrt{2\hbar\omega_b}}\xi_b   
    \end{aligned} 
    \end{cases}
\end{equation}
We analyze Eqs.~(\ref{eq:lang_time}) in the frequency domain, assuming that the initial condition is set in the distant past, \textit{i.e.}, $t_0 \to -\infty$. The aforementioned assumption, combined with the constraint that $f(t)$ is a highly localized function around the zero of its argument (a Dirac delta function for an ohmic bath), allows the last term in \eqref{eqlengev} to be safely neglected. We define the decay rate function $\gamma_j(\omega)$ as the Fourier transform of $\theta(t) f_j(t)$, where $\theta(t)$ is the Heaviside step function, needed to ensures causality. Although in general $\gamma_j(\omega)$ is a complex-valued function, it must satisfy the property $\gamma_j^*(\omega ) = \gamma_j(-\omega )$, since $f_j(t)$ is real-valued function. For the purposes of this paper, we assume $\gamma_j(\omega)$ itself to be real. Nonetheless, in explicit calculations, there is no restriction against taking it as complex. 
Therefore, the resulting quantum Langevin equations for the normal phase in the frequency domain are
\begin{equation}\label{Pippo}
 -i \omega \!\begin{pmatrix}
\tilde{a} \\
\,\;\tilde{a}^{\vphantom{\dagger} \dagger} \\
\tilde{b} \\
\,\;\tilde{b}^{\vphantom{\dagger} \dagger}
\end{pmatrix} \!= -i\! \left({\bf A_{\rm NP}} - \frac{i}{2} {\bf \Gamma}_{\rm NP} \right)\! \!\begin{pmatrix}
\tilde{a} \\
\,\;\tilde{a}^{\vphantom{\dagger} \dagger} \\
\tilde{b} \\
\,\;\tilde{b}^{\vphantom{\dagger} \dagger}
\end{pmatrix}\!  + \tilde{{\bf F}}_{\rm in}\,,
\end{equation}
where we defined the Langevin forces vector \(\bf {\tilde{F}}_{\rm in}\) as
\begin{equation}
    {\bf \tilde{F}}_{\rm in}= \frac{i}{\sqrt{2\hbar}}\left(\!\frac{1}{\sqrt{\omega_a}}\tilde{\xi}_a,\, -\frac{1}{\sqrt{\omega_a}}\tilde{\xi}_a,\, \frac{1}{\sqrt{\omega_b}}\tilde{\xi}_b, \, -\frac{1}{\sqrt{\omega_b}}\tilde{\xi}_b \! \right)^T
\end{equation}
\eqref{Pippo} coincides with \eqref{eq:Langevin_eq_input} in the main text.

\section{Minimization of the Total Hamiltonian in the Superradiant Phase} \label{sec:minimization_H_superradiant}

This Appendix details the procedure outlined in Sec.~\ref{sec:superradiant_phase} for calculating the condensates in the superradiant phase. This approach is based on the minimization of the total system-bath Hamiltonian, ensuring a self-consistent treatment that also accounts for the bath's degrees of freedom. This represents a fundamental distinction from previous theoretical methods, which typically assume a weak system-bath coupling and rely on corresponding approximations.
We point out that we carry out all the following calculations in the thermodynamic limit, where only the states belonging to the maximum angular momentum manifold were retained. This assumption is well-justified, as the states with $j < N/2$ do not contribute to the emission spectra, since they couple more weakly to the bosonic mode (see, e.g., \cite{Zappala2025}).

As already pointed out in the main text, to accurately describe the superradiant phase, it is essential to account for the macroscopic coherent occupation of both system fields and the eventual induced macroscopic occupation that could arise in the bath. Thus, the bosonic system operators have to be shifted as $a = a_{\rm s} + \sqrt{\alpha} $ and $ b = b_{\rm s} - \sqrt{\beta}$, and correspondingly the bath operators as $c_{an} = c_{{\rm s}, an} + \sqrt{\sigma_{an}}$ and $c_{bn} = c_{{\rm s}, bn} - \sqrt{\sigma_{bn}}$ (see \eqref{eq:macroscopic_occupations} of the main text).
The system coordinates, to which the bath operators couple, in the superradiant phase are $ X_a = \sqrt{\hbar/2\omega_a} ( a_{\rm s}^\dagger + a_{\rm s} + 2\sqrt{\alpha})$ and
$X_b = \sqrt{\hbar(N-\beta)/2\omega_a N}\, ( b_{\rm s}^\dagger\sqrt{\theta} + \sqrt{\theta} \, b_{\rm s} - 2\sqrt{\beta}\sqrt{\theta} )$ for systems $A$ and $B$, respectively, where $\theta = 1-[b_{\rm s}^\dagger b_{\rm s} - \sqrt{\beta} (b_{\rm s}^\dagger + b_{\rm s})] (N-\beta)^{-1}$. The additional terms in the definition of $X_b$ in the superradiant phase arise from the non-negligible terms in the expansion of $\sqrt{1-b^\dagger b / N}$, due to the macroscopic occupations.

Upon substituting the aforementioned definitions into the total Hamiltonian in \eqref{eq:H_tot}, and considering the thermodynamic limit, we obtain the total Hamiltonian in terms of the shifted bosonic operators (up to constant terms, here excluded for simplicity)
\begin{widetext}
\begin{equation} \label{eq:Htot_sr}
    \begin{aligned}
    \frac{H}{\hbar} =&\,\, \omega_a a_{\rm s}^\dagger a_{\rm s} + \left[2g \sqrt{\frac{\alpha \beta}{N(N-\beta)}}+\omega_b  \right] b_{\rm s}^\dagger b_{\rm s}  - \left[2g\sqrt{\frac{\beta (N-\beta)}{N}} - \omega_a \sqrt{\alpha} \right] (a_{\rm s}^\dagger+a_{\rm s}) \\
    + & \left[2g(N-2\beta)\sqrt{\frac{\alpha}{N (N-\beta)}} -\omega_b\sqrt{\beta}  \right] (b_{\rm s}^\dagger+b_{\rm s}) + \frac{g(2N-\beta)}{2(N-\beta)}\sqrt{\frac{\alpha \beta}{N(N-\beta)}}\left(b_{\rm s}^\dagger + b_{\rm s} \right)^2 \\
    +&\,\, g(N-2\beta)\sqrt{\frac{1}{N(N-\beta)}}(a_{\rm s}^\dagger+a_{\rm s}) (b_{\rm s}^\dagger+b_{\rm s})\\
    +&\sum_n \! \left\{ \omega_{an} c_{{\rm s}, an}^\dagger c_{{\rm s}, an} - \frac{1}{2} \sqrt{\frac{k_{an} \omega_{an}}{\omega_a}} \left( c_{{\rm s}, an}^\dagger + c_{{\rm s}, an} \right) \left( a_{\rm s}^\dagger + a_{\rm s} \right) + \frac{k_{an}}{4 \omega_a} \left( a_{\rm s}^\dagger + a_{\rm s} \right)^2 \right.\\
    &\qquad + \left. \sqrt{\frac{k_{an}}{\omega_a}} \left( \sqrt{\frac{k_{an} \alpha}{\omega_a}} - \sqrt{\omega_{an} \sigma_{an}} \right) \left( a_{\rm s}^\dagger + a_{\rm s} \right) + \sqrt{\omega_{an}} \left( \sqrt{\omega_{an} \sigma_{an}} - \sqrt{\frac{k_{an} \alpha}{\omega_a}} \right) \left( c_{{\rm s}, an}^\dagger + c_{{\rm s}, an} \right) \right\}\\
    +&\sum_n \! \left\{ \omega_{bn} c_{{\rm s}, bn}^\dagger c_{{\rm s}, bn} - \frac{1}{2} \sqrt{\frac{\tilde{k}_{bn} \omega_{bn}}{\omega_b}} \left( 1 - \frac{\beta}{N - \beta} \right) \!\! \left( c_{{\rm s}, bn}^\dagger + c_{{\rm s}, bn} \right) \!\! \left( b_{\rm s}^\dagger + b_{\rm s} \right) + \frac{\beta}{N - \beta} \! \left( \sqrt{\frac{\tilde{k}_{bn} \omega_{bn} \sigma_{bn}}{\beta \, \omega_b}} - \frac{\tilde{k}_{bn}}{\omega_b} \right) \! b_{\rm s}^\dagger b_{\rm s} \right. \\
    & \qquad \left. + \left[ \frac{1}{2 (N - \beta)} \sqrt{\frac{\tilde{k}_{bn} \omega_{bn} \sigma_{bn} \beta}{\omega_b}} \left( 1 + \frac{\beta}{2(N - \beta)} \right) + \frac{\tilde{k}_{bn}}{4 \omega_b} \left( 1 - \frac{4 \beta}{N - \beta} \right) \right] \left( b_{\rm s}^\dagger + b_{\rm s} \right)^2 \right. \\
    & \qquad + \left. \sqrt{\beta} \left(\! 1 \!- \frac{\beta}{N - \beta} \right) \!\! \left( \sqrt{\frac{\tilde{k}_{bn} \omega_{bn} \sigma_{bn}}{\beta \, \omega_b}} - \frac{\tilde{k}_{bn}}{\omega_b} \right) \!\! \left( b_{\rm s}^\dagger + b_{\rm s} \right) - \sqrt{\omega_{bn}} \left( \! \sqrt{\omega_{bn} \sigma_{bn}} \!- \sqrt{\frac{\tilde{k}_{bn} \beta}{\omega_b}} \right) \!\! \left( c_{{\rm s}, bn}^\dagger + c_{{\rm s}, bn} \right) \!\! \right\}
    \end{aligned}
\end{equation}
\end{widetext}
where $\tilde{k}_{bn} = k_{bn} (\lambda + 1) / 2 \lambda$, with $\lambda \!=\! 4 g^2/(\omega_a \omega_b)$.

We now proceed to the minimization of this Hamiltonian, which corresponds to the vanishing of the linear terms in the bosonic operators in \eqref{eq:Htot_sr}, leading to the non-zero macroscopic occupations
\begin{equation} \label{eq:app_macroscopic_occupations}
    \begin{aligned}
        \alpha & = \frac{N g^2}{\omega_a^2} \left( 1 - \frac{1}{\lambda^2} \right) \,,\quad\quad
        \beta = \frac{N}{2} \left( 1 - \frac{1}{\lambda} \right), \\
        \sigma_{an} & =  \frac{k_{an}}{\omega_{an} \omega_a}\alpha \,,\quad\quad\quad\quad\;\;
        \sigma_{bn} =  \frac{\tilde{k}_{bn}}{\omega_{bn} \omega_b} \beta \, ,
    \end{aligned}
\end{equation}
as in \eqref{eq:macroscopic_occupations} of the main text. 
Inserting these equilibrium values into \eqref{eq:Htot_sr}, the effective total Hamiltonian for the superradiant phase becomes
\begin{equation} \label{eq:H_minimizzata}
    H = H_{\rm SP} + \frac{1}{2} \sum_{j=a,b} \sum_{n} \left[p_{jn}^2 + \tilde{k}_{jn} \left( q_{{\rm s},jn} - x_j \right)^2 \right] \, ,
\end{equation}
coinciding with \eqref{eq:H_superradiant} in the main text.

\section{Quantum Langevin Equations in the Superradiant Phase} \label{sec:quant_lang_eq_superradiant}

To calculate the quantum Langevin equation in the superradiant phase, we apply the same methodology outlined in Sec.~\ref{sec:quant_lang_eq_normal}, since the Hamiltonians in Eqs.~(\ref{eq:H_tot}) and (\ref{eq:H_superradiant}) are formally equivalent, with the only difference being that the bare bosonic operators are replaced by the shifted ones. 
Hence, we obtain
\begin{equation} \label{eq:lang_sr}
\begin{cases}
\begin{aligned}
    \dot{a}_{\rm s} & = -i\omega_a a_{\rm s} -i\tilde{g}(b_{\rm s}^\dagger+b_{\rm s}) \\
    &+\frac{1}{2}\int_{t_0}^t \! f_a(t-t')(a_{\rm s}^\dagger(t')-a_{\rm s}(t'))dt' +\frac{i}{\sqrt{2\hbar\omega_a}}\xi_a \\
    \rule{0pt}{2ex}
    \dot{a}_{\rm s}^\dagger & = i\omega_a a_{\rm s}^\dagger +i\tilde{g}(b_{\rm s}^\dagger+b_{\rm s}) \\
    &-\frac{1}{2}\int_{t_0}^t \! f_a(t-t')(a_{\rm s}^\dagger(t')-a_{\rm s}(t'))dt' -\frac{i}{\sqrt{2\hbar\omega_a}}\xi_a \\
    \rule{0pt}{2ex}
    \dot{b}_{\rm s} & = -i\tilde{\omega}_b b_{\rm s} -i\tilde{g}(a_{\rm s}^\dagger+a_{\rm s})-2iD(b_{\rm s}^\dagger+b_{\rm s}) \\
    & + \frac{1}{2} \frac{2}{\lambda(\lambda+1)}\int_{t_0}^t \! f_b(t-t')(b_{\rm s}^\dagger(t')-b_{\rm s}(t'))dt' \\
    & +\frac{i}{\sqrt{2\hbar\tilde{\omega}_b}}\sqrt{\frac{2}{\lambda(\lambda+1)}} \, \xi_b \\
    \rule{0pt}{2ex}
    \dot{b}_{\rm s}^\dagger & = i\tilde{\omega}_b b_{\rm s}^\dagger +i\tilde{g}(a_{\rm s}^\dagger+a_{\rm s})+2iD(b_{\rm s}^\dagger+b_{\rm s})  \\
    & -\frac{1}{2} \frac{2}{\lambda(\lambda+1)} \int_{t_0}^t \! f_b(t-t')(b_{\rm s}^\dagger(t')-b_{\rm s}(t'))dt' \\
    & -\frac{i}{\sqrt{2\hbar\tilde{\omega}_b}} \sqrt{\frac{2}{\lambda(\lambda+1)}} \, \xi_b
    \end{aligned}
    \end{cases}
\end{equation}
where \(\tilde{\omega}_b\), \(\tilde{g}\) and $D$ are defined in Sec.~\ref{sec:standard_Dicke}. The preceding system of equations can be reformulated within the frequency domain as
\begin{equation}
    -i \omega \!\begin{pmatrix}
        \tilde{a}_{\rm s} \\
        \tilde{a}_{\rm s}^{\vphantom{\dagger} \dagger} \\
        \tilde{b}_{\rm s} \\
        \tilde{b}_{\rm s}^{\vphantom{\dagger} \dagger}
    \end{pmatrix} \!\!=\! -i\! \left( \!{\bf A_{\rm SP}} - \frac{i}{2} {\bf \Gamma}_{\rm SP} \!\right)\! \!\begin{pmatrix}
        \tilde{a}_{\rm s} \\
        \tilde{a}_{\rm s}^{\vphantom{\dagger} \dagger} \\
        \tilde{b}_{\rm s} \\
        \tilde{b}_{\rm s}^{\vphantom{\dagger} \dagger}
    \end{pmatrix}\!  + \! \begin{pmatrix}
         \frac{i}{\sqrt{2\hbar\omega_a}}\tilde{\xi}_a \\
         \frac{-i}{\sqrt{2\hbar\omega_a}}\tilde{\xi}_a \\
         \frac{i}{\sqrt{\hbar\tilde{\omega}_b \lambda (\lambda + 1)}}\tilde{\xi}_b \\
         \frac{-i}{\sqrt{\hbar\tilde{\omega}_b \lambda (\lambda + 1)}}\tilde{\xi}_b
    \end{pmatrix}   
\end{equation}
coinciding with \eqref{eq:Langevin_eq_input} of the main text.

\section{Input-Output Operators} \label{sec:app_inout_op} 

The importance of appropriately defining input-output fields lies in their ability to enable the analytical computation of reflection and transmission spectra in both ohmic and non-ohmic environments. While this aspect has previously been examined in the main text, here we focus on deriving the relationship between Langevin forces, input/output operators and baths degrees of freedom.
We start by performing the continuum limit on \(\xi_j(t)\) in \eqref{eq:xi_ab}, which reads
\begin{align}
     \xi_{j}(t) = \int_0^\infty \sqrt{\frac{\hbar \omega k_j(\omega)}{2}} \left( \vphantom{e^{-i \omega (t-t_0)}} \right. & \tilde{c}_{j}(\omega, t_0) e^{-i \omega (t-t_0)}  \nonumber \\
    & \left. +\, \tilde{c}_{j}^\dagger(\omega, t_0) e^{i \omega (t-t_0)} \right) \,d\omega \, .
\end{align}
Such equation bears a strong resemblance to the derivative of the input operators in \eqref{eq:C_in}, upon the application of the unitary transformation \( \tilde{c}_{j} \to i \tilde{c}_{j} \) to the bath bosonic operators, \textit{i.e.},
\begin{align}
    \dot{C}_{{\rm in}, j}(t) = \int_0^\infty \sqrt{\frac{\hbar\omega}{4 \pi}} \left( \vphantom{e^{-i \omega (t-t_0)}} \right. & \tilde{c}_{j}(\omega, t_0) e^{-i \omega (t-t_0)}  \nonumber \\
    & \left. +\, \tilde{c}_{j}^\dagger(\omega, t_0) e^{i \omega (t-t_0)} \right) \,d\omega \, .
\end{align}
By the microscopic definition of the damping rate, it can be shown that \( k_j(\omega) =2 \gamma_j(\omega)/\pi \), from which the relation in the frequency-domain \( \tilde{\xi}_j(\omega) = -2i\omega \sqrt{\gamma_j(\omega)} \, \tilde{C}_{{\rm in},j}(\omega) \) follows directly.
For the sake of convenience, we explicitly report the expression of \( \tilde{\xi}_j(\omega) \), which is
\begin{align} \label{eq:xi_tilde}
        \tilde{\xi}_j(\omega) =&\, \frac{1}{\sqrt{2\pi}}\int_{-\infty}^\infty\xi_j(t) e^{i\omega t}dt =  \nonumber \\
        =& \int_0^{\!\,\infty} \!\!\! \sqrt{\hbar \pi \omega^\prime k_j(\omega^\prime)}  \left[ \vphantom{\tilde{c}^\dagger_{{\rm in}, j}}\tilde{c}_{{\rm in}, j}(\omega^\prime) \delta(\omega - \omega^\prime)\right. \nonumber \\
        &\left. \quad\qquad\qquad\qquad +\, \tilde{c}^\dagger_{{\rm in}, j}(\omega^\prime) \delta(\omega + \omega^\prime)\right] d\omega^\prime\!\, ,  
\end{align} 
where $\tilde{c}_{{\rm in}, j} (\omega) = \tilde{c}_{j}(\omega, t_0) e^{i \omega t_0}$ is the $j$-th bath bosonic operator evaluated in the remote past, ideally $t_0 \to -\infty$.
The corresponding output operators, $\tilde{c}_{{\rm out}, j}(\omega) = \tilde{c}_{j}(\omega, t_1) e^{i \omega t_1}$, are defined similarly with the only distinction that the bath operators are evaluated in the distant future, $t_1 \to \infty$.
As a direct consequence of \eqref{eq:xi_tilde}, it follows that the Dirac delta determines the selection of annihilation or creation operators for the \( j \)-th bath, depending on the sign of \( \omega \). Specifically, for \( \omega > 0 \), we obtain
\begin{equation}
    \tilde{\xi}_j(\omega) = \sqrt{\hbar \pi \omega k_j(\omega)} \, \tilde{c}_{{\rm in}, j}(\omega),
\end{equation}
whereas for \( \omega < 0 \) the creation operator is selected.
It is now straightforward to establish the connection between the Langevin forces and the input operators in Fourier space 
\begin{equation}
    \tilde{C}_{{\rm in},j}(\omega)=\pm\sqrt{\frac{\hbar \omega_j}{2\gamma_j}}\frac{\tilde{F}_{{\rm in},j}(\omega)}{\omega} \, ,
\end{equation}
where the $+ \, (-)$ is associated to the annihilation (creation) system operators.

\section{Two-mode squeezing} \label{sec:Two mode sqeezing}

In Sec.~\ref{sec:squeezing}, we have thoroughly examined the impossibility of observing intrinsic squeezing in the Dicke model's ground state through electro-optical detection. Our analysis has been primarily focused on the dispersive regime, aiming to highlight the physical aspects of the problem while avoiding unnecessary mathematical complications. 
In this Appendix, we investigate the same topic without resorting to any approximation.

Let us consider the scenario in which the measurement system is placed outside the reference system. Suppose we can extract and analyze the output signals from both the resonator (\(\tilde{C}_{\rm out,a}\)) and the material (\(\tilde{C}_{\rm out,b}\)). We then proceed by following the same steps outlined in Secs.~\ref{sec:theory_spectra} and \ref{sec:squeezing}. 
In accordance with the procedure described in Ref.~\cite{hayashida2023perfect}, for a two-mode squeezing, we consider a general superposition of the two output operators in terms of the angles $\theta$ and $\psi$ as
\begin{equation}\label{eq:E1}
    \tilde{C}_{\theta,\psi}(\omega)=\tilde{C}_{{\rm out}, a}(\omega) \cos{\theta}+e^{i\psi}\tilde{C}_{{\rm out}, b}(\omega) \sin{\theta} \,.
\end{equation}

Similarly to \eqref{eq:X+_X-}, we define the positive and negative frequency operators \(X_{\theta,\psi}^+(\omega)\) and \(X_{\theta,\psi}^-(\omega)\), respectively, by
\begin{equation}
    \begin{aligned}
        X_{\theta,\psi}^+(\omega)&= \tilde{C}_{\theta,\psi}(\omega) \, ,\\
        X_{\theta,\psi}^-(\omega)&= \tilde{C}^\dagger_{\theta,\psi}(\omega)\, .
    \end{aligned}
\end{equation}
Therefore, the quadrature operator is \(X_{\phi,\theta,\psi}(\omega)=e^{i\phi}X_{\theta,\psi}^+(\omega)+e^{-i\phi}X_{\theta,\psi}^-(\omega)\), from which we evaluate its ground-state variance \(
(\Delta X_{\phi,\theta,\psi})^2 \equiv \bra{0}(X_{\phi,\theta,\psi})^2\ket{0} - \bra{0}X_{\phi,\theta,\psi}\ket{0}^2 = \bra{0}(X_{\phi,\theta,\psi})^2\ket{0}
\),
where \(\ket{0}\) is the ground state. Taking into account the relationship between the output and input operators as given in \eqref{eq:output_vs_input}, along with their representation in terms of the bath operators in \eqref{eq:inbath}, we obtain the following result
\begin{equation}
      (\Delta X_{\phi,\theta,\psi})^2 = \bra{0} X_{\theta,\psi}^+(\omega)X_{\theta,\psi}^-(\omega)\ket{0}\,.
\end{equation}
As previously observed in the main text, the fact that \((\Delta X_{\phi,\theta,\psi})^2\) is independent of the angle \(\phi\) implies that squeezing in the ground state cannot be detected using these techniques.

To recover the results obtained in Sec.~\ref{sec:squeezing} within the dispersive regime (\(\omega_b \gg \omega_a\)), we set \(\theta = 0\) to reduce the problem to single-mode squeezing. This leads to 
\begin{equation}
      (\Delta X_{\phi,0,\psi})^2 = \bra{0} X_{0,\psi}^+(\omega)X_{0,\psi}^-(\omega)\ket{0}=\frac{\hbar}{2\omega}\, .
\end{equation}

\bibliographystyle{plain}

\end{document}